\documentclass[aps,pra,onecolumn,showpacs,notitlepage,superscriptaddress,letterpaper]{revtex4-1}
\usepackage{graphicx}
\usepackage{amsmath}
\usepackage{esint}
\usepackage{verbatim}
\usepackage{color}
\usepackage{SIunits}
\usepackage{hyperref}

\newcommand{\nbar}{\langle n \rangle}
\newcommand{\nbath}{n_{\text{b}}}
\newcommand{\nbathp}{n_{\text{p}}}
\newcommand{\Tbathp}{T_{\text{p}}}
\newcommand{\Tbath}{T_{\text{b}}}
\newcommand{\Tf}{T_{\text{f}}}

\newcommand{\lambdacO}{\lambda_{\text{c}}}
\newcommand{\ncavO}{n_{\text{c}}}
\newcommand{\gzeroO}{g_{\text{0}}}

\newcommand{\kappai}{\kappa_{\text{i}}}
\newcommand{\kappae}{\kappa_{\text{e}}}

\newcommand{\gammapO}{\gamma_{\text{p}}}
\newcommand{\gammanotO}{\gamma_{\text{0}}}
\newcommand{\gammabO}{\gamma_\text{b}}
\newcommand{\omegacO}{\omega_{\text{c}}}
\newcommand{\omegamO}{\omega_{\text{m}}}
\newcommand{\gammaOMO}{\gamma_{\text{OM}}}

\newcommand{\QmO}{Q_{\text{m}}}

\newcommand{\omegalO}{\omega_{\text{l}}}
\newcommand{\nNEP}{n_\text{NEP}}

\newcommand{\gammaSB}{\Gamma_\text{SB,0}}
\newcommand{\gammaPump}{\Gamma_\text{pump}}
\newcommand{\gammaDark}{\Gamma_\text{dark}}
\newcommand{\gammaTot}{\Gamma_\text{tot}}
\newcommand{\etaSPD}{\eta_\text{SPD}}

\newcommand{\Tper}{T_\text{per}}
\newcommand{\Tpulse}{T_\text{pulse}}
\newcommand{\tpulse}{t_\text{pulse}}
\newcommand{\ncavon}{n_\text{c,on}}
\newcommand{\ncavoff}{n_\text{c,off}}
\newcommand{\deltab}{\delta_\text{b}}
\newcommand{\gammaS}{\gamma_\text{S}}
\newcommand{\Ceff}{C_\text{eff}}
\newcommand{\nH}{n_\text{H}}
\newcommand{\nD}{n_\delta}
\newcommand{\ahat}{\hat{a}}
\newcommand{\adag}{\hat{a}^{\dagger}}
\newcommand{\bhat}{\hat{b}}
\newcommand{\bdag}{\hat{b}^{\dagger}}
\newcommand{\ain}{\hat{a}_\text{in}}
\newcommand{\aindag}{\hat{a}^{\dagger}_\text{in}}
\newcommand{\bin}{\hat{b}_\text{in}}
\newcommand{\bindag}{\hat{b}^{\dagger}_\text{in}}
\newcommand{\aout}{\hat{a}_\text{out}}
\newcommand{\Ain}{\hat{A}_\text{in}}
\newcommand{\Aindag}{\hat{A}_\text{in}^{\dagger}}
\newcommand{\Aout}{\hat{A}_\text{out}}
\newcommand{\Aoutdag}{\hat{A}_\text{out}^{\dagger}}
\newcommand{\Bin}{\hat{B}_\text{in}}
\newcommand{\Bindag}{\hat{B}_\text{in}^{\dagger}}
\newcommand{\Bout}{\hat{B}_\text{out}}

\newcommand{\Fhat}{\hat{F}}
\newcommand{\Fdag}{\hat{F}^{\dag}}
\newcommand{\Fcorr}{\left<\Fdag\Fhat\right>}
\newcommand{\Uhat}{\hat{U}}
\newcommand{\Udag}{\hat{U}^{\dagger}}
\newcommand{\UhatO}{\hat{U}_0}
\newcommand{\UdagO}{\hat{U}^{\dagger}_0}
\newcommand{\Uhatm}{\hat{U}_\text{m}}
\newcommand{\Udagm}{\hat{U}^{\dagger}_\text{m}}

\newcommand{\aphi}{\hat{a}_\phi}
\newcommand{\aphidag}{\hat{a}^\dagger_\phi}
\newcommand{\Spp}{S_{\phi\phi}}
\newcommand{\Saa}{S_{\alpha\alpha}}
\newcommand{\omegaf}{\omega_\text{f}}
\newcommand{\kappaf}{\kappa_\text{f}}
\newcommand{\alphain}{\alpha_\text{in}}
\newcommand{\alphaout}{\alpha_\text{out}}
\newcommand{\af}{\hat{a}_\text{filt}}
\newcommand{\afdag}{\hat{a}^\dagger_\text{filt}}
\newcommand{\ai}{\hat{a}_\text{i}}
\newcommand{\Sbb}{S_\text{bb}}

\begin{document}

%\title{Optothermal dynamics of a nanomechanical resonator in its quantum ground state of motion}
%\title{Optical back-action, heating, and damping dynamics of a nanomechanical resonator at milliKelvin temperatures}
%\title{Transient dynamics of a nanomechanical resonator near its quantum ground-state of motion}
\title{Pulsed excitation dynamics of an optomechanical crystal resonator near its quantum ground-state of motion}

\author{Se\'{a}n M.\ Meenehan}
%\thanks{These authors contributed equally to this work.}
\author{Justin D.\ Cohen}
%\thanks{These authors contributed equally to this work.}
\author{Gregory S. MacCabe}
%\thanks{These authors contributed equally to this work.}
\affiliation{Kavli Nanoscience Institute and Thomas J. Watson, Sr., Laboratory of Applied Physics, California Institute of Technology, Pasadena, CA 91125, USA}
\affiliation{Institute for Quantum Information and Matter, California Institute of Technology, Pasadena, CA 91125, USA}
\author{Francesco Marsili}
\author{Matthew D. Shaw}
\affiliation{Jet Propulsion Laboratory, California Institute of Technology, Pasadena, CA 91109, USA}
\author{Oskar Painter}
\email{opainter@caltech.edu}
\affiliation{Kavli Nanoscience Institute and Thomas J. Watson, Sr., Laboratory of Applied Physics, California Institute of Technology, Pasadena, CA 91125, USA}
\affiliation{Institute for Quantum Information and Matter, California Institute of Technology, Pasadena, CA 91125, USA}  

\date{\today}
\begin{abstract}
Using pulsed optical excitation and read-out along with single phonon counting techniques, we measure the transient back-action, heating, and damping dynamics of a nanoscale silicon optomechanical crystal cavity mounted in a dilution refrigerator at a base temperature of $\Tf \approx 11$~mK. In addition to observing a slow ($\sim740$~ns) turn-on time for the optical-absorption-induced hot phonon bath, we measure for the $5.6$~GHz `breathing' acoustic mode of the cavity an initial phonon occupancy as low as $\nbar = 0.021 \pm 0.007$ ($\Tbath \approx 70$~mK) and an intrinsic mechanical decay rate of $\gammanotO = 328 \pm 14$~Hz ($\QmO \approx 1.7\times10^7$). These measurements demonstrate the feasibility of using short pulsed measurements for a variety of quantum optomechanical applications despite the presence of steady-state optical heating.
\end{abstract}
\maketitle

The recent cooling of nanomechanical resonators to their motional quantum ground state~\cite{OConnell2010,Teufel2011b,Chan2011} opens the possibility of utilizing engineered mechanical systems strongly coupled to optical or microwave fields for a variety of quantum metrology and information processing applications~\cite{Aspelmeyer2014}, amongst them the preparation of highly non-classical mechanical states~\cite{Vanner2013,Borkje2011,Galland2014} and coherent frequency conversion between microwave and optical signals~\cite{Stannigel2010,Safavi-Naeini2011a,Hill2012,Bochmann2013,Andrews2014}.  A particularly interesting device architecture for realizing large radiation pressure coupling between light and mechanics is the thin-film optomechanical crystal (OMC)~\cite{Eichenfield2009b,Safavi-Naeini2010b}, in which optical and acoustic waves can be guided and co-localized via patterning of the surface layer of a microchip.  Based largely upon the OMC concept, new ideas for phononic quantum networks~\cite{Habraken2012} and optomechanical metamaterials~\cite{Schmidt2013} have been proposed, in which arrays of cavity-optomechanical resonators are coupled together via optical or acoustic degrees of freedom, and in which laser light is used to parametrically control the emergent network or material properties.         

For many of the above mentioned applications, operation at cryogenic temperatures is desired as it offers a route to obtaining the requisite low thermal occupancy and long mechanical coherence time.  Recent measurements at millikelvin (mK) bath temperatures of an OMC resonator formed from single crystal silicon~\cite{Eichenfield2009b,Chan2012}, however, have shown substantial mechanical mode heating and mechanical damping due to weak sub-bandgap optical absorption~\cite{Meenehan2014}.  Although optical $Q$-factors in excess of $10^6$ are realized in these highly optimized structures~\cite{Chan2012}, the large impact of even very weak optical absorption can be attributed to a combination of the relatively large energy per photon, and the sharp drop in thermal (phonon) conductance with temperature in the low temperature limit.  Further complications arise from the seemingly contradictory requirements of isolating the mechanical resonator from its environment to obtain high mechanical $Q$-factor, and that of providing large thermal anchoring to a low temperature bath for cooling of the mechanical resonator.  

%In the work of Ref.~\cite{Meenehan2014} we partially overcame this issue by employing a phononic bandgap material to connect the mechanical resonator to the low temperature bath, thus allowing high frequency ($>7GHz)$ phonons to carry heat from the resonator, while isolating from its environment the mechanical mode of interest in the phononic bandgap.            

In this work we utilize pulsed optical excitation and single phonon counting~\cite{Cohen2014} to study the transient dynamics of optical back-action, heating, and damping of the $5.6$~GHz mechanical mode of a silicon optomechanical crystal resonator at mK bath temperatures.  Phonon counting, realized by photon counting of the optically filtered motional sidebands of the reflected optical excitation pulse, yields simultaneously a high time resolution ($\sim 25$~ns) and mechanical mode occupancy sensitivity ($<10^{-2}$).  Measurement of both Stokes and anti-Stokes sidebands also yields an absolute calibration of the occupancy of the resonator mode in terms of mechanical vacuum noise~\cite{Safavi-Naeini2012,Purdy2014,Weinstein2014}.  In addition to measuring initial phonon mode occupancies as low as $\nbar = 0.021 \pm 0.007$ and mechanical decay times as long as $\tau = 475 \pm 21$~$\mu$s, we observe a slow ($\sim740$~ns) turn-on time for the optical-absorption-induced phonon bath that both heats and damps the mechanical resonator mode.  Taken together, these measurements demonstrate the feasibility of using short pulsed measurements for quantum optical state engineering of the mechanics in silicon optomechanical crystals, despite the presence of large steady-state optical heating.
\begin{figure}[btp]
\begin{center}
\includegraphics[width=\textwidth]{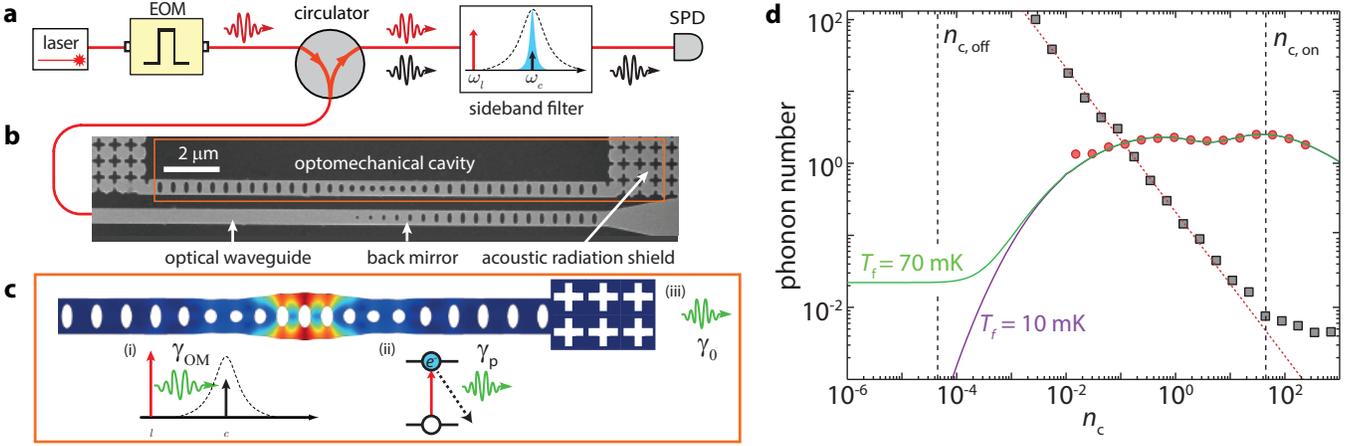}
\caption{\textbf{a}, Pulsed pump light at frequency $\omegalO$ (red arrows) is directed to an optomechanical crystal cavity (OMC) inside a dilution fridge via an optical circulator. The cavity reflection is then filtered at the cavity frequency $\omegacO$ (black arrows) and directed to a single photon detector (SPD). \textbf{b}, SEM image of the silicon OMC cavity studied in this work.  \textbf{c},  Finite-element-method simulation of the localized acoustic resonance at $5.6$~GHz of the OMC cavity.  Deformation of the beam structure is exaggerated to highlight the mechanical motion, with color indicating the regions of high (red) and low (blue) displacement magnitude.  The inset shows the three processes affecting the mode occupancy: (i) detuned pump light at $\omegalO$ (red) exchanges energy with the acoustic mode at rate $\gammaOMO$, generating scattered photons at $\omegacO$ (black) in the process, (ii) pump light drives excitation of electronic defect states in the silicon device layer, subsequently exciting a hot phonon bath which heats the localized acoustic resonance at rate $\gammapO$, and (iii) phonons escape the cavity volume via the acoustic shield at intrinsic decay rate $\gammanotO$, coupling the localized resonance to the fridge bath. \textbf{d}, Noise-equivalent phonon number $\nNEP$ (gray squares) and phonon occupancy $\nbar$ (red circles) for red-detuned ($\Delta = \omegamO$) continuous-wave (CW) pumping versus intracavity photon number $\ncavO$. The red dashed line indicates the expected $\nNEP$ contribution from SPD dark counts. The vertical dashed line at $\ncavO \approx 45$ $(4.5\times10^{-5})$ indicates the photon number during the on-state (off-state) of the pulse. Solid green and purple lines show fits to the CW heating model for fridge bath temperatures of $\Tf = 70$~mK and $10$~mK, respectively.} \label{fig:setup_and_sensitivity}
\end{center} 
\end{figure}

The device studied in this work consists of a patterned nanobeam, formed from the top silicon device layer of a silicon-on-insulator wafer. The etched hole pattern in the silicon nanobeam forms an optomechanical crystal, in which photonic and phononic bandgaps at the ends of the beam co-localize optical and acoustic resonances with frequencies of $\omegacO/2\pi \approx 196$~THz (free space wavelength $\lambdacO \approx 1530$~nm) and $\omegamO/2\pi \approx 5.6$~GHz, respectively. A scanning electron micrograph (SEM) of the device is shown in Fig.~\ref{fig:setup_and_sensitivity}b, and a finite-element method simulation of the acoustic resonance is displayed in Fig.~\ref{fig:setup_and_sensitivity}c. Coupling to the optical resonance is accomplished via an end-fire coupling scheme, using a lensed optical fiber inside of a dilution refrigerator to couple to the central waveguide shown in the SEM, as described in Ref.~\cite{Meenehan2014}. Surrounding the cavity is a 2D cross pattern~\cite{Safavi-Naeini2010b} which possesses a complete acoustic bandgap in the $5-6$~GHz range, providing an additional acoustic shield for the mechanical resonator mode while allowing phonons above and below the phononic bandgap to carry heat from the nanobeam structure.

The experimental set-up is shown schematically Fig.~\ref{fig:setup_and_sensitivity}a. Electro-optic modulation of a laser probe beam generates optical pulses with frequency and duty-cycle controlled by a variable delay electrical pulse generator. The red arrows indicate a coherent pump at frequency $\omega_{\text{l}}$ which is red-detuned from the optical resonance frequency $\omegacO$ by $\Delta \equiv \omegacO - \omega_{\text{l}} = \omegamO$. In this case absorption of a single phonon from the nanomechanical resonator results in upconversion of a pump photon to the anti-Stokes sideband at $\omegacO$, represented by the black arrows. The cavity reflection is filtered to reject the pump frequency and subsequently directed to a high-efficiency single-photon detector (SPD) and a time-correlated single photon counting system synced to the pulse generator. This measurement repeats each pulse period, building up a histogram with respect to photon arrival time relative to the sync pulse during a certain integration time. As the number of anti-Stokes photons is directly proportional to the average phonon occupancy of the mechanical resonator, $\nbar$, the photon count rate in each time bin then portrays the time-evolution of $\nbar(t)$ during the pulse on-state. All measurements presented herein were performed at a fridge base temperature of $\Tf = 11$~mK. Further details about device fabrication and the experimental setup can be found in Refs.~\cite{Chan2012,Cohen2014} and in the appendices.

The signal-to-noise ratio (SNR) of this phonon counting method is determined by the sideband scattering rate $\gammaOMO \equiv 4 \gzeroO^2 \ncavO/\kappa$ ($\gzeroO$ is the optomechanical coupling rate, $\ncavO$ is the intracavity photon number, and $\kappa$ is the optical decay rate), the total system detection efficiency ($\eta$), the dark count rate of the SPD ($\gammaDark$), and the residual transmission of the filters at the pump frequency relative to the peak transmission ($A$). A useful parameterization of the sensitivity to low $\nbar$ is the noise-equivalent phonon occupancy $\nNEP$, defined as~\cite{Cohen2014}
\begin{equation}
\nNEP = \frac{\gammaDark}{\eta \gammaOMO} + A \left(\frac{\kappa \omegamO}{2 \kappae \gzeroO}\right)^2,
\end{equation}
\noindent where $\kappae$ is the optical decay rate into the detection channel. For the device under test we have $\kappa/2\pi = 443$~MHz, $\kappae/2\pi = 221$~MHz, and $\gzeroO/2\pi = 710$~kHz. A typical measured $\nNEP$ for our setup, taken using a comparable device at room temperature, is shown in Fig.~\ref{fig:setup_and_sensitivity}d as gray squares. Here, to obtain the lowest $\nNEP$ optical pre-filtering is used to remove phase noise around the laser line and broadband spontaneous emission from the optical probe beam, details of which are given in App.~\ref{sec:appE}.  For sufficiently high probe power ($\ncavO>40$), $\nNEP$ falls below $10^{-2}$, enabling sensitive detection of the mechanical resonator deep in its quantum ground state. However, at sub-kelvin temperatures optical absorption heating produces a steady-state $\nbar>1$ for $\ncavO>0.01$ (red circles in Fig.~\ref{fig:setup_and_sensitivity}c) during continuous-wave (CW) optical excitation. In order to maintain the OMC in the mechanical ground state, the duty cycle of the pulse train must be kept sufficiently low, and the modulation depth sufficiently high, such that the mechanical mode thermalizes to the dilution refrigerator ambient bath between successive pulses.

The CW behavior fits well to a thermal model consisting of the three processes illustrated in Fig~\ref{fig:setup_and_sensitivity}c: (i) the radiation pressure coupling at rate $\gammaOMO$ between the mechanical mode and the effective zero-temperature probe laser, (ii) coupling to an optical-absorption-induced hot phonon bath above the phononic shield bandgap, and (iii) coupling to the ambient fridge bath at an intrinsic rate $\gammanotO$ through the acoustic radiation shield.  At the low intra-cavity photon numbers of these measurements, we believe the optical absorption heating is a result of excitation of electronic defect states at the silicon surfaces~\cite{Stesmans1996,Borselli2006}, and subsequent phonon-assisted relaxation of these states.  As detailed in Ref.~\cite{Meenehan2014}, the resulting local hot-phonon bath occupancy ($\nbathp$) is found to scale as $\ncavO^{1/4}$ in steady-state, consistent with linear optical absorption and a cubic drop in the thermal conductance with temperature~\cite{Holland1963}.  The corresponding coupling rate of the mechanical resonance to the high frequency hot-phonon bath ($\gammapO$) is measured to scale as $\Tbathp \exp(-\hbar\omega_{\text{c}}/k_{\text{B}}\Tbathp)$ for low bath temperature ($\Tbathp < 4$~K), corresponding to inelastic phonon scattering with a quasi-equilibrium hot phonon bath above a cut-off phonon frequency of $\omega_{\text{c}}/2\pi \approx 35$~GHz.  As shown in Fig~\ref{fig:setup_and_sensitivity}d, extrapolation of this steady-state model for a fridge base temperature of $\Tf=10$~mK reveals a relevant pulse-off-state regime of $\ncavO<10^{-4}$ in which absorption heating effects are negligible.
\begin{figure}[btp]
\begin{center}
\includegraphics[width=0.5\textwidth]{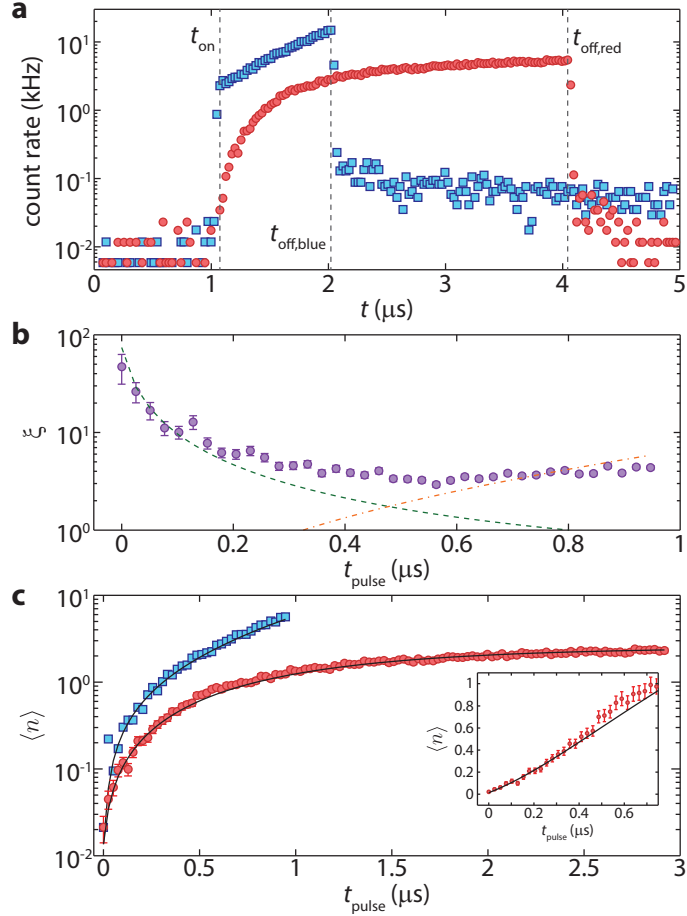}
\caption{\textbf{a} Total photon count rate versus time $t$ for a red-detuned (red circles, $\Delta = \omegamO$) and blue-detuned (blue squares, $\Delta = \omegamO$). For both data sets $\ncavon \approx 45$ and $\Tper = 5$~ms. Vertical dashed lines indicate the start and stop times of the pulses. \textbf{b}, Asymmetry $\xi$ versus time within the pulse $\tpulse$. The green dashed and orange dash-dotted lines show the theoretically expected $\xi (t)$ in the absence of backaction and heating effects, respectively. Error bars show one standard deviation (s.d.) determined from the measured count rates, assuming Poissonian counting statistics. \textbf{c}, Calibrated phonon occupancy $\nbar$ versus $\tpulse$ for $\Delta = \omegamO$ (red circles) and $\Delta = -\omegamO$ (blue squares). The solid black lines show fits to a model including a slow exponential turn-on of the hot phonon bath. The inset shows detail for $\tpulse < 750$~ns on a linear scale for $\Delta = \omegamO$. Error bars show one s.d. determined from the measured count rates, assuming Poissonian counting statistics. For the blue-detuned data the error bars are smaller than the data markers.} \label{fig:pulse_and_asym}
\end{center} 
\end{figure}

Figure~\ref{fig:pulse_and_asym}a shows the measured sideband photon count rate versus time with pulsed optical excitation, for both red- ($\Delta = \omegamO$) and blue-detuned ($\Delta = -\omegamO$) pumping. The on-state pulse amplitude in these measurements is $\ncavon=45$, corresponding to an optomechanical coupling rate of $\gammaOMO/2\pi = 205$~kHz.  For the achievable pulse extinction ratio of $60-70\;$dB, this results in a residual off-state photon number of $\ncavoff<4.5\times10^{-5}$.  Vertical dashed lines indicate the time bins corresponding to the start and stop of the pulse, determined from observing the rising and falling edges of the pulse when bypassing the cavity. The start and stop times are taken to be the time bins for which the pulse reaches $90\%$ of its maximum value. Henceforth, the variable $t$ refers to time relative to the synchronization signal generated by the pulse generator, while $\tpulse$ refers to time relative to the start of the optical pulse occurring around $t \approx 1 ~\mu$s.  The pulse period in these measurements is fixed at $\Tper = 5$~ms.  

Throughout the pulse a pronounced asymmetry is observed between count rates for red-detuned versus blue-detuned pumping, which can be quantified by the asymmetry parameter $\xi = \Gamma_{-}/\Gamma_{+}-1$, where $\Gamma_{\pm}$ is the sideband photon count rate for a pump detuning $\Delta = \pm\omegamO$.  While motional sideband asymmetry has previously been measured in a variety of opto- and electromechanical systems using linear detection schemes~\cite{Safavi-Naeini2012,Weinstein2014,Lee2014,Purdy2014} , the use of phonon counting techniques allows the observed asymmetry to be directly and unambiguously attributed to quantum fluctuations of the mechanical oscillator~\cite{Khalili2012,Weinstein2014}.  This asymmetry, shown versus $\tpulse$ in Fig.~\ref{fig:pulse_and_asym}b, initially decreases with time before leveling off and beginning to increase for sufficiently long pulse times. The increase at later times can be ascribed to the effect of optomechanical backaction, which results in cooling or heating of the mechanical resonator for red- or blue-detuned pumps, respectively~\cite{Kippenberg2007}. However, for $\tpulse \ll \gammaOMO^{-1} \approx 750$~ns the effects of backaction can be neglected and the phonon occupancy $\nbar$ may be assumed equal for both pump detunings. In this case, the asymmetry is simply related to the occupancy by $\xi = \nbar^{-1}$, and arises from the fundamental asymmetry between phonon absorption ($\Gamma_{+} \propto \nbar$) and emission ($\Gamma_{-} \propto \nbar+1$) processes~\cite{Diedrich1989,Safavi-Naeini2012}. Theoretical plots of the two contributions to $\xi (t)$ are shown in Fig.~\ref{fig:pulse_and_asym}b. The green dashed line shows the expected $\xi (t)$ assuming that $\nbar\left(\Delta =-\omegamO\right) = \nbar\left(\Delta =\omegamO\right)$ (no backaction), while the orange dash-dotted line shows $\xi (t)$ in the absence of optical heating and in the case when $\nbath \gg 1$ such that $\xi (0) = 0$ and asymmetry arises solely from backaction effects. From the asymmetry measured in the first 25 ns time bin, we extract a minimum phonon occupancy of $\nbar_\text{min} = 0.021 \pm 0.007$ ($T_\text{min} \approx 70$~mK).  Extrapolation to zero time yields a mode temperature closer to $40$~mK, but still above the $11$~mK of the mixing plate of the fridge, likely indicating a higher local chip temperature.  

%This measured occupancy is lower than previous results in both cavity optomechanical~\cite{Chan2011} and electromechanical~\cite{Teufel2011b} systems by more than an order of magnitude, and is comparable to the occupancy measured in other GHz frequency mechanical oscillators at mK temperatures~\cite{OConnell2010}. 

\begin{figure}[btp]
\begin{center}
\includegraphics[width=0.5\textwidth]{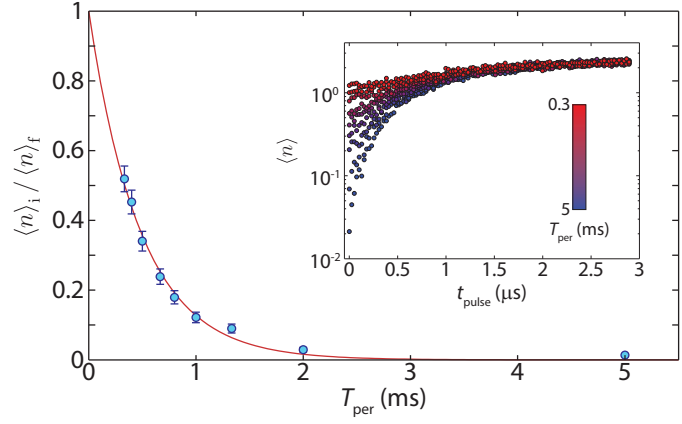}
\caption{Ratio of initial/final phonon number $\nbar_\text{i}/\nbar_\text{f}$ versus pulse period $\Tper$. The solid red curve shows a fit to exponential decay. Error bars show one s.d. determined from the measured count rates, assuming Poissonian counting statistics The inset shows $\nbar$ versus $\tpulse$ for the different values of $\Tper$.} \label{fig:ringdown}
\end{center} 
\end{figure}

Once $\nbar$ is determined in the initial time bin it may be used to convert measured count rates to phonon occupancies throughout the pulse for either pump detuning. The resulting calibrated occupancy versus $\tpulse$ is shown in Fig.~\ref{fig:pulse_and_asym}c. A simple model of the heating dynamics would suggest an exponential increase towards the steady-state phonon occupancy with total rate $\gamma = \gammanotO+\gammapO+\gammaOMO$. However, as can be seen in the inset to Fig.~\ref{fig:pulse_and_asym}c, the curvature of the occupancy curve is positive for short times. This fact is inconsistent with the simple model described above, and is likely due to a finite equilibration time of the hot phonon bath. The data can be fit well for both detunings by using a simple model which assumes that a small fraction of the hot phonon bath turns on effectively instantaneously (i.e. very fast relative to the length of a single time bin) while the remainder has a slow exponential increase to its steady-state value. Thus, the effective rate equation for the phonon occupancy is given by
\begin{equation}
\dot{\nbar} = -\gamma \nbar +\gammapO \nbathp \left(1-\deltab e^{-\gammaS \tpulse}\right),
\end{equation}
\noindent where $\deltab$ and $\gammaS$ are the slow growing fraction of the hot phonon bath and the corresponding turn-on rate, respectively, and the steady-state hot phonon bath parameters $\gammapO$ and $\nbathp$ can be determined by fitting the CW $\nbar$ versus $\ncavO$ data shown in Fig.~\ref{fig:setup_and_sensitivity}d. Fitting the pulsed occupancy curve yields $\deltab = 0.79\pm0.08$ and $\gammaS/2\pi = 215\pm29$~kHz, indicating that the heating occurs slowly enough as to be manageable during coherent quantum operations as determined in the remaining discussion.

During the off-state of the pulse, optical heating of the mechanical resonator should be negligible and the phonon occupancy should cool at the intrinsic damping rate $\gammanotO$. Thus, the initial and final occupancies during the pulse ($\nbar_\text{i}$ and $\nbar_\text{f}$, respectively) should obey the relation $\nbar_\text{i} = e^{-\gammanotO \Tper}\nbar_\text{f}$, assuming the pulse period $\Tper$ is much larger than the pulse width $\Tpulse$. The ratio $\nbar_\text{i}/\nbar_\text{f}$, shown in Fig.~\ref{fig:ringdown}a versus $\Tper$, displays the expected exponential decay with $\gammanotO/2\pi = 328 \pm 14$~Hz and a corresponding intrinsic mechanical quality factor of $\QmO \approx 1.7 \times 10^7$. This decay rate corresponds well with the value inferred from previous CW measurements of occupancy at mK temperatures~\cite{Meenehan2014}. From our measurement of $\gammanotO$ and $\nbar_\text{min}$ we estimate a thermal decoherence time of $\tau_\text{th} = (\gammanotO(1+\nbar_\text{min}))^{-1} = 475 \pm 21~\mu$s.
\begin{figure}[btp]
\begin{center}
\includegraphics[width=0.5\textwidth]{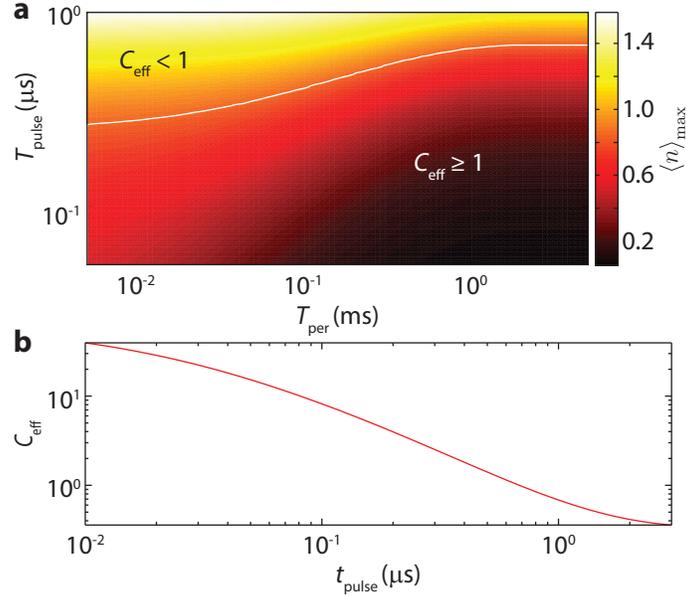}
\caption{\textbf{a}, Maximum phonon occupancy during a pulse, $\nbar_\text{max}$, versus pulse period $\Tper$ and pulse width $\Tpulse$. The white contour delineates the region where the effective cooperativity $\Ceff \geq 1$ throughout the pulse. \textbf{b}, $\Ceff$ versus time within the pulse $\tpulse$ for $\Tper = 5$~ms, $\Tpulse = 3$~$\mu$s.} \label{fig:nmax_and_ceff}
\end{center} 
\end{figure}

While the low thermal occupancy and long coherence time measured here are promising, the utility of cavity optomechanical systems for performing coherent quantum operations between the optical and mechanical degrees of freedom is ultimately predicated upon the ability to simultaneously achieve $\nbar \ll 1$ and large cooperativity $C \equiv \gammaOMO/\gammabO$, where $\gammabO = \gammanotO+\gammapO$ is the total coupling rate between the mechanical resonator and its thermal bath. In the specific example of optomechanically mediated coherent transfer of photons between optical and superconducting microwave resonators~\cite{Stannigel2010,Safavi-Naeini2011a}, the relevant figure of merit is the effective cooperativity $\Ceff \equiv C/\nbath$, where $\nbath$ is the effective bath occupancy defined such that $\gammabO \nbath = \gammanotO n_0 + \gammapO \nbathp$, which must be much larger than unity in order to achieve low-noise photon conversion at the single quantum level~\cite{Hill2012,Andrews2014}. Using the measured values of $\gammanotO$ and $\gammaS$, we can calculate the maximum phonon occupancy achieved during the pulse, $\nbar_\text{max}$, for a given $\Tpulse$ and $\Tper$ and $\Delta = \omegamO$  (Fig.~\ref{fig:nmax_and_ceff}a), as well as $\Ceff$ as a function of $\tpulse$ (Fig.~\ref{fig:nmax_and_ceff}b).

Though optical heating of our devices prevents us from reaching the $\Ceff >1$ regime using a CW pump, due to the slow turn-on of the hot phonon bath observed in this work we find that $\nbar < 1$ and $\Ceff >1$ can be maintained throughout the entire pulse period for $\Tpulse \lesssim 300$~ns at pulse rates approaching $1$~MHz. Furthermore, we find $\Ceff \gg 1$ during the initial $100$~ns of the pulse and reaches values as large as $\Ceff \approx 40$. An analysis of the effects of optical heating on pulsed phonon addition and subtraction~\cite{Vanner2013} for the creation of non-Gaussian mechanical states are presented in App.~\ref{sec:appG}, wherein we calculate the fidelity of generating a mechanical Fock state~\cite{Galland2014} to be $F > 0.8$ for pulse rates up to 1 kHz. These calculations demonstrate that optomechanical applications in the quantum regime are feasible with our current structure for reasonable pulse parameters without further mitigation of optical heating effects.

\begin{acknowledgments}
The authors would like to thank V. B. Verma, R. P. Miriam and S. W. Nam for their help with the single photon detectors used in this work.  This work was supported by the DARPA ORCHID and MESO programs, the Institute for Quantum Information and Matter, an NSF Physics Frontiers Center with support of the Gordon and Betty Moore Foundation, the AFOSR through the ``Wiring Quantum Networks with Mechanical Transducers'' MURI program, and the Kavli Nanoscience Institute at Caltech.  Part of the research was carried out at the Jet Propulsion Laboratory, California Institute of Technology, under a contract with the National Aeronautics and Space Administration.
\end{acknowledgments}

%\bibliography{../Mirror_v2}

\appendix

\section{Device Fabrication}
\label{sec:appA}
 The devices are fabricated from a silicon-on-insulator (SOI) wafer (SOITEC, 220 nm device layer, 3 $\mu$m buried oxide) using electron beam lithography followed by reactive ion etching (RIE/ICP). The Si device layer is then masked using ProTEK PSB photoresist to define a mesa region of the chip to which a tapered lensed fiber can access. Outside of the protected mesa region, the buried oxide is removed with a plasma etch and a trench is formed in the underlying silicon substrate using tetramethylammonium hydroxide (TMAH). The devices are then released in hydrofluoric acid ($49~\%$ aqueous HF solution) and cleaned in a piranha solution (3-to-1 H$_2$SO$_4$:H$_2$O$_2$) before a final hydrogen termination in diluted HF. In fabrication, arrays of the nominal design are scaled by $\pm2~\%$ to account for frequency shifts due to fabrication imperfections and disorder.

\section{Experimental Setup}
\label{sec:appB}

\begin{figure}[btp]
\begin{center}
\includegraphics[width=\textwidth]{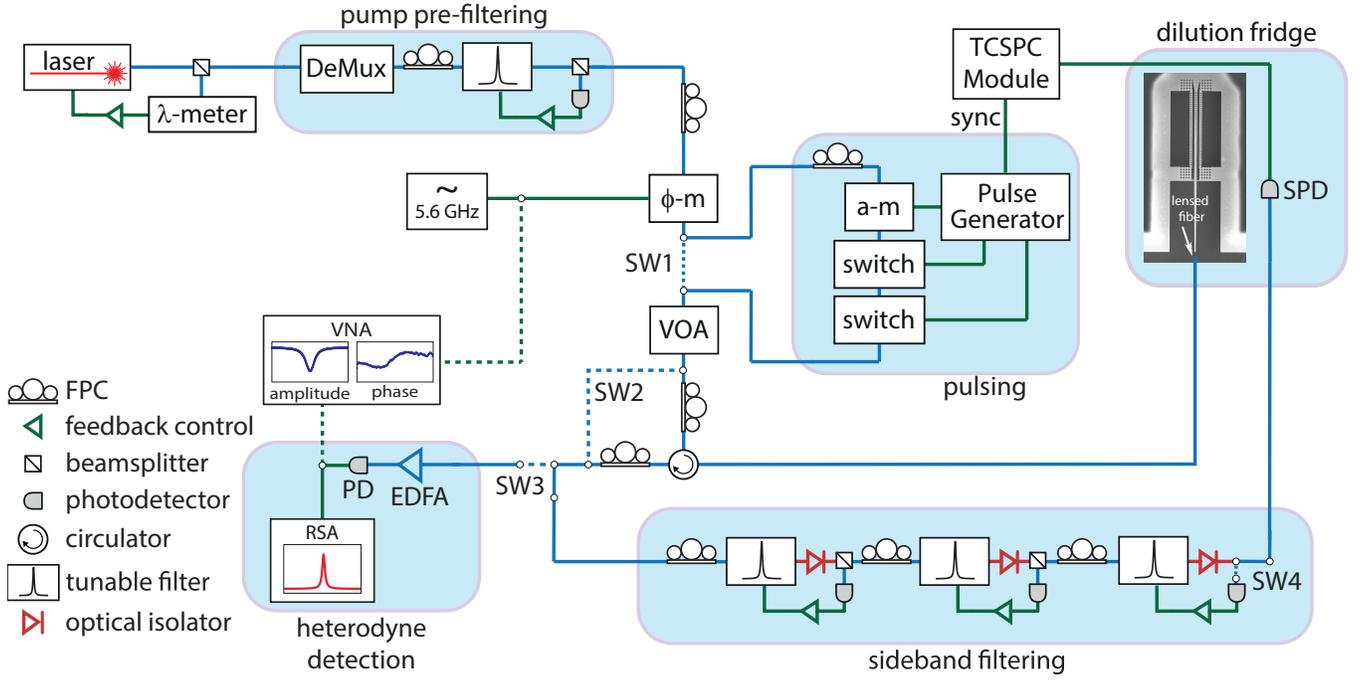}
\caption{\textbf{Experimental setup.} Full setup for performing pulsed phonon counting. $\lambda$-meter: wavemeter, DeMux: C-band optical demultiplexer, $\phi$-m: electro-optic phase modulator, SW: optical switch, a-m: high-extinction electro-optic amplitude modulator, switch: high-speed electro-optic switch, VOA: variable optical attenuator, EDFA: erbium-doped fiber amplifier, PD: high-speed photodetector, RSA: real-time spectrum analyzer, VNA: vector network analyzer, SPD: single photon detector, TCSPC: time-correlated single photon counting, FPC: fiber polarization controller..} \label{fig:setup_SI}
\end{center} 
\end{figure}

The full measurement setup is shown in Fig.~\ref{fig:setup_SI}. A fiber-coupled, wavelength tunable external cavity diode laser is used as the light source, with a small portion ($\sim 1\%$) of the laser output sent to a wavemeter ($\lambda$-meter) for frequency stabilization. The remaining laser power is sent through both a C-band optical demultiplexer (DeMux) to reject broadband spontaneous emission from the laser, and a high-finesse tunable filter to remove laser phase noise at the mechanical frequency, which can both contribute to excess pump transmission noise on the single photon detector~\cite{Cohen2014}. After pre-filtering, the laser pump is sent through an electro-optic phase modulator ($\phi$-m) which may be driven by either an RF signal generator, to generate optical sidebands for locking the filter cavities, or a vector network analyzer (VNA) for obtaining the amplitude and phase response of the optical cavity. 

To generate pulses, an optical switch (SW1) allows the pump tone to be sent through a series of electro-optic amplitude modulators for the generation of optical pulses. The first is a fast high-extinction modulator (a-m) with a rise and fall time of $\sim25$~ns. Though in principle the high-extinction modulator can provide $50-60$~dB extinction on its own, its transmission level is much less stable and difficult to lock at the maximum extinction point. Thus, we lock the modulator at $\sim30$~dB of extinction and use two electro-optic switches, each providing $\sim18 - 20$~dB extinction, to fully extinguish the pump. While these switches are much slower ($\sim 200$~ns rise time, $\sim30$~$\mu$s fall time), the extra switching time should have a negligible impact as it is much smaller than the pulse periods used here ($\sim1$~ms or more). The modulators are collectively driven by a variable delay electrical pulse generator which also provides a synchronization pulse to the single photon counting electronics. 

A variable optical attenuator (VOA) controls the power input to the cavity, after which an optical circulator routes the laser light to a lensed fiber tip inside the dilution refrigerator for end-fire coupling to the device. Subsequently, the cavity reflection can be switched (SW3) into one of two detection setups. The first setup sends the signal through an erbium-doped fiber amplifier (EDFA) followed by a high-speed photodetector (PD). The resulting amplified photocurrent may be directed to a real-time spectrum analyzer (RSA) in order to measure the optical noise power spectral density (NPSD) for mechanical characterization or to the VNA which is used in conjunction with the phase modulator to measure the full complex response of the optical cavity for purposes of optical characterization as described below. The second detection path sends the reflected signal through three additional tunable filters in order to reject the pump frequency, and then back into the dilution refrigerator where it is  detected by a superconducting single photon detector (SPD). The output of this detector is sent to a time-correlated single photon counting (TCSPC) module to build up a histogram of photon count events as a function of time relative to the sync pulse received from the pulse generator.

The SPDs used in this work are amorphous WSi-based superconducting nanowire single-photon detectors developed in collaboration between the Jet Propulsion Laboratory and NIST. The SPDs are design for high-efficiency detection of individual photons in the wavelength range $\lambda = 1520-1610$~nm with maximum count rates of about $2.5\times 10^7$ counts per second (c.p.s)~\cite{Marsili2013}. The SPDs are mounted on the still stage of the dilution refrigerator at $\sim700$~mK. Single-mode optical fibers (Corning SMF-28) are passed into the refrigerator through vacuum feedthroughs and coupled to the SPDs via a fiber sleeve attached to each SPD mount. Proper alignment of the incoming fiber with the 15~$\mu$m~$\times$~15~$\mu$m square area of the SPD nanowire is ensured by a self-aligned mounting system incorporated into the design of the SPD~\cite{Marsili2013}. The radio-frequency output of each SPD is amplified by a cold-amplifier mounted on the 50 K stage of the refrigerator as well as a room-temperature amplifier, then read out by a triggered PicoQuant PicoHarp 300 time-correlated single photon counting module. By systematically isolating the input optical fiber from environmental light sources and filtering out long wavelength blackbody radiation inside the fridge we have achieved dark count rates of $\sim4$~(c.p.s.). At just below the switching current of the detectors, we have measured a peak detection efficiency of $\etaSPD = 68\%$, with $\lesssim 20\%$ variability depending on photon polarization.

The tunable filters used for both pre-filtering the pump and filtering the cavity reflection are commercially available, piezo-tunable Fabry-Perot filters (Micron Optics, FFP-TF2), all with a $\sim50$~MHz bandwidth, a free-spectral range of $\sim20$~GHz, and a tuning voltage of $\lesssim18$~V per free-spectral range. The filters each offer roughly $40$~dB of pump suppression (relative to the peak transmission) measured at a filter-pump varies by $1-2$~dB from filter to filter. When locking the post-cavity filters a switch is used to bypass the cavity (SW2), as a relatively large amount of CW power is used during the locking procedure and we would like to avoid sending large amounts of power into the cavity unless necessary. More importantly, the SPD is also bypassed (SW4), as allowing too much power to reach the SPD will saturate it resulting in significantly elevated dark counts ($\gammaDark \approx 500-1000$ c.p.s.) for $1-2$~minutes after the signal is turned off. Once the switches are set an RF signal generator is used to drive the phase modulator at $\omegamO$, producing an optical sideband which is aligned with the cavity (and motional sideband) frequency. To lock the filter chain, a sinusoidal voltage (with an initial range of $\pm10$~V) is used to dither each filter while monitoring its transmission. The offsets of the sinusoids are then adjusted, and their amplitudes reduced, to maximize transmission of the desired sideband, fixing the voltages once all three filters are well aligned. Over time the filters will drift and the locking procedure will need to be repeated, though in subsequent re-locking attempts a much smaller dithering amplitude of $\sim1$~V is sufficient. As the piezo elements controlling each filter have a finite settling time, the filters will drift much more rapidly following the initial locking attempt (during which large voltage shifts are applied) than following subsequent re-locks. After several re-locks, the filters will typically become stable enough that the total transmission at the sideband frequency changes by $\lesssim5\%$ over several minutes. At this point the phase modulator is turned off, the pulse generator, cavity and SPD are switched back into the optical train and we begin accumulating pulsed data for $2$~minutes before re-locking the filters. The total filter transmission is recorded at the end of a locking procedure and again prior to re-locking, and if the transmission has shifted by more than a few percent the previous pulsed dataset is discarded. All the time-resolved data shown in this chapter are taken in this manner, with all $2$~minutes datasets averaged together to produce the final pulsed histogram. 

\section{Device Characterization}
\label{sec:appC}
The measurements presented in this work rely on an accurate calibration of the optomechanical damping rate $\gammaOMO$, which depends on the vacuum optomechanical coupling rate $\gzeroO$, the total optical decay rate $\kappa$, and the intracavity photon number $\ncavO$. The photon number for a given power and detuning in turn depends on the single pass fiber-to-waveguide coupling efficiency $\eta_\text{cpl}$ and waveguide-cavity coupling efficiency $\eta_\kappa = \kappae/\kappa$ ($\kappae$ is the waveguide-cavity coupling rate). The fiber collection efficiency is determined by measuring the calibrated reflection level far-off resonance with the optical cavity on the optical power meter, and is found to be $\eta_\text{cpl} = 0.68$ for the device measured here (total device reflection efficiency of $\sim0.46\%$). To measure $\kappa$ and $\eta_\kappa$, the laser is placed off-resonance from the cavity and the VNA is used to drive the phase modulator and sweep an optical sideband across the cavity. By detecting the reflection on a high-speed photodiode connected to the VNA input we obtain the amplitude and phase response of the cavity from which can extract $\kappa/2\pi = 443$~MHz (corresponding to an optical quality factor of $Q_\text{o} \approx 4.4\times10^5$), and $\eta_\kappa = 0.5$. With these three parameters measured, it is possible to determine $\ncavO$ for an arbitrary input power to the cavity.

To characterize the acoustic resonance, the EDFA is used to amplify the cavity reflection so that the optical noise floor overcomes the photodetector's electronic noise and the optical NPSD is measured on the RSA, where a Lorentzian response due to transduction of the acoustic thermal Brownian motion can be observed at the acoustic resonance frequency $\omegamO/2\pi = 5.6$~GHz. For a red- or blue detuned pump laser ($\Delta \equiv \omegacO-\omegalO = \pm \omegamO$) the linewidth of this transduced noise peak is $\gamma = \gammabO \pm \gammaOMO$, where $\gammabO$ is the bare coupling rate of the mechanical resonator to its bath and $\gammaOMO \equiv 4\gzeroO^2\ncavO/\kappa$. Thus, by averaging the observed linewidth for red- and blue-detuned pumps we can determine $\gammabO$ and thus extract the excess optomechanically induced damping rate $\gammaOMO$ as a function of $\ncavO$. With $\ncavO$ and $\kappa$ determined, a linear fit of $\gammaOMO$ reveals a vacuum optomechanical coupling rate of $\gzeroO/2\pi = 710$~kHz.

\section{Phonon Counting}
\label{sec:appD}
In the linearized sideband resolved regime with a red-detuned pump ($\Delta = \omegamO$) the total reflected cavity amplitude will be sum of the coherent pump reflection $\alphaout$ and a fluctuation term given, in the Fourier domain and in a frame rotating at the pump frequency, by~\cite{Safavi-Naeini2013a}
\begin{equation}
\aout(\omega)|_{\Delta=\omegamO} \approx r(\omega)\ain(\omega)+n(\omega)\ai(\omega)+s(\omega)\bin(\omega), \label{eqn:aFourier}
\end{equation}

\noindent where $\ain$ and $\ai$ are the pump noise and intrinsic cavity noise, respectively (assumed for the moment to be vacuum noise), $\bin$ is the noise operator for the mechanical bath, which has the correlation function $\left<\bindag(\omega)\bdag(\omega')\right> = \nbath\delta(\omega+\omega')$ ($\nbath$ is the bath occupancy), and the effective scattering matrix elements are given by
\begin{gather}
r(\omega) = 1-\frac{2\kappae}{\kappa}\pm\frac{\gammaOMO\kappae}{\kappa}\frac{1}{\pm i\left(\omegamO\mp\omega\right)+\gamma/2}, \\
n(\omega;\pm) = \pm\frac{\sqrt{\kappai\kappae}}{\kappa}\left(\frac{\gammaOMO}{\pm i\left(\omegamO\mp\omega\right)+\gamma/2}\mp 2\right), \\ 
s(\omega) = \sqrt{\frac{\kappae}{\kappa}}\frac{i\sqrt{\gamma_\text{i}\gammaOMO}}{\pm i\left(\omegamO\mp\omega\right)+\gamma/2}.
\end{gather} 

As the full reflected signal from the cavity includes the reflected pump amplitude, which is many orders of magnitude larger than the sideband signal, we must first filter the cavity output to reject the pump. Provided the filter is sufficiently high-finesse, it can be modeled in the frequency domain by a single Lorentzian function
\begin{equation}
F(\omega;\omegaf) = \frac{\kappaf/2}{i(\omegaf-\omega)+\kappaf/2}, \label{eqn:filterDef}
\end{equation}

\noindent where $\kappaf$ and $\omegaf$ are the resonant frequency of the filter, respectively. Now, explicitly considering the case of red-detuned driving in the sideband-resolved regime, the filtered cavity output will be the product of $F(\omega;\omegaf)$ with the frequency domain output of the cavity. As the resonantly enhanced anti-Stokes sideband photons will be detuned by $\omegamO$ from the pump, we choose $\omegaf = \omegamO$. Using Eq.~\ref{eqn:aFourier}, we obtain
\begin{equation}
\af(\omega) = F(\omega;\omegamO)\left(\alphaout\delta(\omega)+r(\omega;+)\ain(\omega)+n(\omega;+)\ai(\omega)+s(\omega;+)\bin(\omega)\right).
\end{equation}

\noindent Performing photon counting on the filtered output then results in an average count rate of
\begin{align}
\Gamma(t) &= \left<\afdag(t)\af(t)\right> \nonumber \\
&= \frac{1}{2\pi}\int_{-\infty}^{\infty}d\omega\int_{-\infty}^{\infty}d\omega'\; e^{i(\omega+\omega')t} \left<\afdag(\omega)\af(\omega')\right> \nonumber \\
&= \frac{1}{2\pi}\left(|F(0;\omegamO)|^2|\alphaout|^2+\frac{\kappae}{\kappa}\gammaOMO\int_{-\infty}^{\infty}d\omega|F(\omega;\omegamO)|^2\;\Sbb(\omega;\nbar)\right) \nonumber \\
&\approx A|\alphaout^2|+\frac{\kappae}{\kappa}\gammaOMO\nbar,
\end{align}

\noindent where $A = \frac{1}{2\pi}|F(0;\omegamO)|^2$ is the pump attenuation factor, and $\Sbb(\omega;\nbar)$ is the phonon spectral density~\cite{Safavi-Naeini2013a}
\begin{equation}
\Sbb(\omega) = \frac{\gamma\nbar}{(\omegamO+\omega)^2+(\gamma/2)^2}.
\end{equation}

\noindent Note that we have assumed a filter bandwidth $\kappaf \gg \gamma$, allowing us to approximate $|F(\omega;\omegamO)|^2 \approx |F(\omegamO;\omegamO)|^2 = 1$ inside the integral over $\Sbb$. A similar analysis for blue detuning (where $\omegaf = -\omegamO$ to filter the Stokes sideband) yields a comparable result, with $\nbar \rightarrow \nbar+1$. Note that this formula assumes that $\ain$ and $\ai$ are vacuum noise, and thus do not contribute to counting of real photons.

The total count rate, including noise of the photon counter and reduction of the pump and signal due to measurement inefficiency, is given for red-detuning by
\begin{equation}
\gammaTot = \gammaDark + \gammaPump + \eta\frac{\kappae}{\kappa}\gammaOMO\nbar,
\end{equation}

\noindent where $\gammaDark$ is the dark count rate of the photon detector, $\gammaPump = \eta A|\alphaout|^2$ and $\eta$ is the total measurement efficiency. These expressions can be used to perform thermometry in a similar fashion to linear detection, either by calibrating the cavity parameters and total measurement efficiency or by measuring the asymmetry between the red- and blue-detuned count rates.

To assess the sensitivity of this counting scheme, it is convenient to express the measurement noise floor $\gammaDark+\gammaPump$ in terms of an equivalent number of mechanical quanta (that is the mechanical occupancy $\nbar$ that would be needed to yield a signal-to-noise of one). This noise-equivalent phonon number is obtained by dividing the total noise floor by the per-phonon count rate $\gammaSB = \eta(\kappae/\kappa)\gammaOMO$, yielding
\begin{equation}
\nNEP = \frac{\gammaDark+\gammaPump}{\gammaSB}.
\end{equation}

\noindent For a highly sideband-resolved system, the reflected pump in the case of $\Delta = \pm \omegamO$ will be approximately given by $\alphaout \approx \alphain$. This in turn can be expressed in terms of the intracavity photon number as $|\alphaout|^2 \approx \omegamO^2\ncavO/\kappae$. Thus, $\nNEP$ as a function of $\ncavO$ is given by
\begin{equation}
\nNEP(\ncavO) = \frac{\kappa^2\gammaDark}{4\eta\kappae\gzeroO^2\ncavO}+A\left(\frac{\kappa\omegamO}{2\kappae\gzeroO}\right)^2. \label{eqn:nNEP}
\end{equation}

\section{Effects of Technical Laser Noise}
\label{sec:appE} 
The mechanical frequency of the nanobeam used in this experiment ($\omegamO/2\pi = 5.6$~GHz) raises concerns about the effects of laser phase noise on the measurements, as the laser used in this experiment has previously been observed to possess a prominent phase noise peak at $5$~GHz~\cite{Safavi-Naeini2013a}. In addition to phase noise, most diode lasers typically have a small amount of broadband spontaneous emission. While this additional noise is orders of magnitude weaker than the laser tone itself, it exists outside the wavelength region ($\lambda \sim 1520-1570$~nm) where the filters are guaranteed to be high finesse, and thus can be transmitted with high efficiency to the SPDs.

Phase noise in particular is worrisome as it can not only lead to an excess noise floor but also to real heating of the mechanics and systematic errors in thermometry due to noise squashing/anti-squashing~\cite{Poggio2007,Rocheleau2010}. Phase noise can be accounted for by assuming a total pump noise operator given by~\cite{Safavi-Naeini2013a}
\begin{equation}
\ahat_\text{in,tot}(t) = \ain(t)+\aphi(t), \label{eqn:aintotDef}
\end{equation}

\noindent where $\ain(t)$ still represents vacuum noise and
\begin{equation}
\aphi(t) = i|\alphain|\phi(t),
\end{equation}

\noindent where $|\alphain|$ is the amplitude of the pump and $\phi(t)$ is the stochastic phase of the pump, assumed to be a real, stationary Gaussian process with zero mean. The phase noise is assumed to be delta-correlated in the frequency domain, such that
\begin{equation}
\left<\phi(\omega)\phi(\omega')\right> = \Spp(\omega)\delta(\omega+\omega'), \label{eqn:phiCorr}
\end{equation}

\noindent where the expectation value here corresponds to an ensemble average. The phase noise input to the cavity then has the correlation function
\begin{equation}
\left<\aphidag(\omega)\aphi(\omega')\right> = \Saa(\omega)\delta(\omega+\omega'), \label{eqn:aphiCorr}
\end{equation}

\noindent where $\Saa(\omega) = |\alpha|^2\Spp(\omega)$. In a sideband-resolved system, for either red or blue detuning $\Delta = \pm\omegamO$, we find that the presence of phase noise heats the mechanical resonator, with an additional added phonon occupancy given by~\cite{Safavi-Naeini2013a}
\begin{equation}
\nbar_\phi|_{\Delta=\pm\omegamO} = \frac{\kappae}{\kappa}\frac{\gammaOMO}{\gamma}n_\phi,
\end{equation} 

\noindent where we have defined $n_\phi = \Saa(\omegamO)$, and where we have assumed that $\Saa(\omega)$ is sufficiently slow-varying in the vicinity of $\omega = \omegamO$ (specifically for $|\omega-\omegamO| \lesssim \gamma$) that we may approximate $\Saa(\omega) = \Saa(\omegamO)$.

Including this additional noise term in the analysis of the previous section yields a phase noise contribution to the total photon count rate of
\begin{equation}
\Gamma_\phi|_{\Delta=\pm\omegamO} = \eta \int_{-\infty}^{\infty}\frac{d\omega}{2\pi} \Saa(\omega)|F(\omega,\pm\omegamO)|^2|r(\omega;\pm)|^2.
\end{equation}

\noindent If we assume that $\Saa(\omega)$ is slowly-varying in frequency for $|\omega-\omegamO| \lesssim \gamma$ and that $\kappaf \gg \gamma$, this simplifies to
\begin{equation}
\Gamma_\phi|_{\Delta=\pm\omegamO} = \eta n_\phi \left(\frac{\kappaf}{4}\left(1-\frac{2\kappae}{\kappa}\right)^2+\frac{\kappae\gammaOMO}{\kappa}\left(\frac{\gammaOMO\kappae}{\gamma\kappa}\pm\left(1-\frac{2\kappae}{\kappa}\right)\right)\right). \label{eqn:Gammaphi}
\end{equation}

\noindent Using the fact that $n_\phi = |\alpha|^2\Spp(\omegamO) \approx \omegamO^2\ncavO\Spp(\omegamO)/\kappae$, we obtain the contribution of phase noise to the noise-equivalent phonon number
\begin{equation}
n_{\text{NEP},\phi}|_{\Delta=\pm\omegamO} = \left(\frac{\omegamO\kappa}{2\kappae\gzeroO}\right)^2\Spp(\omegamO)\left(\frac{\kappaf}{4}\left(1-\frac{2\kappae}{\kappa}\right)^2+\frac{\kappae\gammaOMO}{\kappa}\left(\frac{\gammaOMO\kappae}{\gamma\kappa}\pm\left(1-\frac{2\kappae}{\kappa}\right)\right)\right). \label{eqn:nNEPphi}
\end{equation}

\noindent Like the pump-bleed through, phase noise leads to a constant contribution to $\nNEP$, and leads to squashing or anti-squashing of the noise depending on detuning and cooperativity, similar to heterodyne detection. Note, however, that in the case $\kappae/\kappa = 0.5$ the contribution of phase noise will not depend upon detuning. Thus, even in the presence of large phase noise it is possible to avoid detuning dependent noise squashing/anti-squashing, though one will still have a large overall phase noise floor. 

It is useful for characterization purposes to calculate the phase-noise contribution to the observed count rates when the laser is far-detuned from the cavity resonance ($|\Delta| \gg \omegamO$). Assuming that the laser-filter detuning is kept constant at $\pm\omegamO$, the phase-noise count rate in this case is just
\begin{align}
\Gamma_\phi|_{|\Delta|\gg\omegamO} &= \eta\int_{-\infty}^{\infty} \frac{d\omega}{2\pi}\Saa(\omega)|F(\omega,\pm\omegamO)|^2 \nonumber \\
&= \eta \frac{\kappaf}{4} n_\phi,
\end{align}

\noindent with a corresponding noise-equivalent phonon number
\begin{equation}
n_{\text{NEP},\phi}|_{|\Delta|\gg\omegamO} = \left(\frac{\omegamO\kappa}{4\kappae\gzeroO}\right)^2\kappaf\Spp(\omegamO). \label{eqn:nNEPphi_detuned}
\end{equation}

To get rid of the excess noise, we insert both the bandpass filter (for filtering spontaneous emission) and a tunable high-finesse filter (for filtering phase noise) immediately after the pump laser output as shown in Fig.~\ref{fig:setup_SI}, enabling us to reach $\nNEP \ll 1$ using a three-filter phonon counting setup as shown in Fig. 1d of the main text. A conservative estimate of the residual phase noise can be made by assuming that the limiting value of $\nNEP \approx 4\times10^{-3}$ is entirely due to phase noise (i.e. perfect filtering of the pump tone). Using Eq.~\ref{eqn:nNEPphi_detuned} we find $\Spp(\omegamO) \lesssim 8\times10^{-19}$~Hz$^{-1}$. For the pump power during the on-state of the pulse ($\ncavon \approx 45$), the corresponding effective phase noise occupancy is $n_\phi \approx 3.2\times10^{-5}$, which has a negligible effect on the measurements in this work.

\section{Heating model}
\label{sec:appF}
The simplest thermal model assumes that the optically induced bath turns on instantaneously when the pulse is in the on-state. The corresponding rate equation for the phonon occupancy $\nbar$, for red- ($\Delta = \omegamO$) and blue-detuned ($\Delta = -\omegamO$) pulses during the on-state is thus
\begin{equation}
\dot{\nbar} = -\gamma \nbar +\gammapO \nbathp + \gammanotO n_0 + \frac{1}{2} \left(1 \mp 1\right) \gammaOMO,
\end{equation}

\noindent where $\gamma = \gammanotO+\gammapO\pm\gammaOMO$, $\gammapO$ and $\nbathp$ are the coupling rate and occupancy of the hot phonon bath, $\gammanotO$ and $n_0$ are the coupling rate and occupancy of the ambient fridge bath, and the extra factor of $\gammaOMO$ for a blue-detuned pump accounts for the possibility of spontaneous emission into the mechanical resonator due to the optomechanical interaction. This rate equation has the simple solution
\begin{align}
\nbar(t) & = \nbar(t_0) e^{-\gamma t} + \nH \left(1-e^{-\gamma t}\right), \\
\nH & = \gamma^{-1}\left(\gammapO \nbathp + \gammanotO n_0 +\frac{1}{2} \left( 1 \mp 1 \right) \gammaOMO\right).
\end{align}

\noindent where $t_0$ is the start time of the pulse, and $t_0 \leq t \leq t_0+\Tpulse$.

In principle, $\gammaOMO$ can be determined independently as described above, $\gammanotO$ can be determined by fitting the occupancy decay during the pulse off-state (Fig. 3 in the main text), while $\gammapO$ and $\nbathp$ can be subsequently determined by fitting the steady-state occupancy curve shown in Fig. 1c of the main text. However, using these independently determined values in a fit to the red- and blue-detuned data shown in Fig. 2a of the main text results in a poor fit and inconsistent results. In particular, the apparent heating rate $\gamma$ is much smaller than expected for a red-detuned pulse and larger than expected for a blue-detuned pulse. This, along with the ``kink" in the heating curve shown in the inset of Fig. 2c of the main text, suggests additional complexity in the heating dynamics.

The simplest addition to the heating model is to assume a finite time for the hot phonon bath to come into equilibrium, which is approximated by allowing a fraction of the hot phonon bath occupancy to turn on exponentially over time. Thus, the phenomenological rate equation becomes
\begin{equation}
\dot{\nbar} = -\gamma \nbar +\gammapO \nbathp \left(1-\deltab e^{-\gammaS t}\right) + \gammanotO n_0 + \frac{1}{2} \left(1 \mp 1\right) \gammaOMO,
\end{equation}

\noindent where $\deltab$ is the slow growing fraction of $\nbathp$ and $\gammaS$ the turn-on rate. Strictly speaking $\gammapO$ should depend on the phonon distribution of the hot phonon bath, and thus would be expected to be time-dependent in this model as well. However, the resulting rate equation becomes intractable in such a case and the effects should be negligible in the regime $\gammaOMO \gg \gammapO$, so we approximate $\gammapO$ equal to its steady-state value. This modified rate equation has the solution
\begin{align}
\nbar(t) = \nbar(t_0) e^{-\gamma t} + \nH \left(1-e^{-\gamma t}\right) + \nD \left( e^{-\gammaS t}-e^{-\gamma t}\right), \; \nD = \frac{\gammapO \nbathp \deltab}{\gammaS-\gamma},
\label{eqn:heat_model}
\end{align}
\noindent which is used to obtain the fit shown in Fig. 2c of the main text with $\nbar(t_0)$, $\gammaS$ and $\deltab$ as free parameters.

During the off-state of the pulse ($t_0+\Tpulse \leq t \leq t_0+\Tper$), the resonator will simply cool towards the ambient fridge occupancy $n_0$ at the intrinsic damping rate $\gammanotO$. Using the initial condition that $\nbar(t_0) = n_0$ for the first pulse ($t_0 = 0$), and iterating many pulses we find that in the steady-state the initial phonon occupancy during a pulse (assuming $\Tper \gg \Tpulse$) is
\begin{equation}
\nbar(t_0) = \frac{n_0 \left(1-e^{-\gammanotO \Tper}\right)+\nH\left(1-e^{-\gamma \Tpulse}\right) e^{-\gammanotO \Tper}+\nD \left(e^{-\gammaS \Tpulse}-e^{-\gammaS \Tper}\right) e^{-\gammanotO \Tper}}{1-e^{-\gamma \Tpulse - \gammanotO \Tper}}.
\label{eqn:n_init}
\end{equation}

Thus, once $n_0$, $\gammaS$ and $\deltab$ are determined by fitting the occupancy curves, we may use Eqs.~\ref{eqn:heat_model} and~\ref{eqn:n_init} to determine the occupancy throughout the pulse for arbitrary $\Tper$ and $\Tpulse$. This allows us to determine the maximum attained phonon occupancy as a function of pulse parameters, as shown in Fig. 4a of the main text.

\section{Phonon addition/subtraction fidelity}
\label{sec:appG}
In general, the optomechanical interaction allows for the creation of a variety of non-classical mechanical states via phonon subtraction or addition using appropriate red- or blue-detuned pulses~\cite{Vanner2013}. In particular, recent proposals have explicitly shown how to herald single phonon Fock states~\cite{Galland2014} and entangled mechanical states~\cite{Borkje2011} via application of a short blue-detuned pulse to an optomechanical system in its motional quantum ground state and subsequent detection of the emitted sideband photon. However, the fidelity of the generated states in our system is in principle degraded by the presence of optical heating during the pulse, which we analyze to lowest order here.

In a frame rotating at the pump frequency, the full Heisenberg-Langevin equations for the optomechanical system are~\cite{Safavi-Naeini2013a}
\begin{align}
\dot{\ahat} & = -\left(i\Delta +\frac{\kappa}{2}\right)\ahat+i\gzeroO\left(\bhat+\bdag\right)\ahat+\sqrt{\kappa}\ain, \\
\dot{\bhat} & = -\left(i\omegamO+\frac{\gammabO}{2}\right)\bhat+i\gzeroO\adag\ahat+\sqrt{\gammabO}\bin,
\end{align}
\noindent where $\ahat$, $\bhat$ are the photon and phonon annihilation operators, respectively, and $\ain$, $\bin$ are quantum noise operators corresponding to the optical and mechanical baths.

We linearize about a classical steady-state by displacing $\ahat \rightarrow \alpha +\ahat$, where $|\alpha|^2 = \ncavO$. For concreteness we will consider the case $\Delta \equiv \omegacO-\omega_\text{l} = -\omegamO$ (blue-detuned pump). Moving into a frame rotating at the mechanical frequency (i.e. $\ahat \rightarrow \ahat e^{i\omegamO t}$, and so on for all operators), and making the rotating wave approximation, valid in the weak coupling ($\gzeroO \sqrt{\ncavO} \ll \kappa$) and sideband-resolved ($\kappa/\omegamO \ll 1$) limit, we obtain
\begin{align}
\dot{\ahat} & = -\frac{\kappa}{2}\ahat+iG\bdag+\sqrt{\kappa}\ain, \\
\dot{\bhat} & = -\frac{\gammabO}{2}\bhat+iG\aindag+\sqrt{\gammabO}\bin,
\end{align}
\noindent where $G = \gzeroO\sqrt{\ncavO}$ and $\ain$, $\bin$ are the usual noise operators multiplied by $e^{-i\omegamO t}$. The noise operators obey the following commutation and correlation relations
\begin{gather}
\left[\ain (t),\aindag (t')\right] = \left[\bin (t), \bindag (t')\right] = \delta(t-t') \\
\left<\ain (t) \aindag(t') \right> = \delta(t-t') \\
\left<\bindag (t) \bin (t') \right> = \nbath (t) \delta(t-t'), \quad \left<\bin (t) \bindag (t') \right> = \left(\nbath (t)+1\right) \delta(t-t').
\end{gather}

Since we are working in the weak-coupling limit ($G \ll \kappa$) we may use the adiabatic solution for $\ahat$ (i.e. $\dot{\ahat} \approx 0$). Moreover, we wish to include the effects of mechanical noise to lowest order. Considering the effect of a short optical pulse of duration $\tau$, we consider the case $\gammabO \tau, \gammabO \int_0^{\tau} ds\;\nbath(s) 
\ll 1$, as well as $\gammaOMO \tau \ll 1$ and $\gammaOMO \gg \gammabO$, where here $\gammaOMO \equiv 4G^2/\kappa$ refers to the \textit{magnitude} of the optomechanical damping rate (the sign will be explicitly incorporated for simplicity, since we're only considering blue-detuning). Under these assumptions, as in Ref.~\cite{Hofer2011}, the mechanical noise term will be retained in the mechanical equation of motion only, since in the optical equations of motion it will acquire an extra factor of $\gammaOMO \tau$ and thus can be neglected to lowest order. Furthermore, the $\left(\gammabO/2\right)\bhat$ term will be neglected in favor of the much larger $\left(\gammaOMO/2\right)\bhat$ term. With these approximations, we arrive at the approximate equations
\begin{align}
\ahat & \approx i \sqrt{\frac{\gammaOMO}{\kappa}} \bdag + \frac{2}{\kappa} \ain, \\
\dot{\bhat} & \approx \frac{\gammaOMO}{2} \bhat +i\sqrt{\gammaOMO}\aindag+\sqrt{\gammabO}\bin. 
\end{align}

Formally integrating the equation of motion for $\bhat$ over the duration of the pulse while using the input-output relation for the optical cavity $\aout = -\ain+\sqrt{\kappa}\ahat$, we arrive at the approximate equations
\begin{gather} 
\aout(t) = \ain(t)+i\sqrt{\gammaOMO} e^{\frac{\gammaOMO t}{2}}\bdag(0) +\gammaOMO e^{\frac{\gammaOMO t}{2}} \int_0^{t} ds\; e^{\frac{-\gammaOMO s}{2}} \ain(s)\\
\bhat(t) = e^{\frac{\gammaOMO t}{2}} \bhat(0)+i\sqrt{\gammaOMO} e^{\frac{\gammaOMO t}{2}} \int_0^{t} ds\; e^{\frac{-\gammaOMO s}{2}} \aindag(s) + \sqrt{\gammabO} e^{\frac{\gammaOMO t}{2}} \int_0^{t} ds\; e^{\frac{-\gammaOMO s}{2}} \bin(s).
\end{gather}

We now introduce the following temporal modes~\cite{Hofer2011}
\begin{gather}
\Bin = \bhat(0), \quad \Bout = \bhat(\tau), \\
\Ain = \sqrt{\frac{\gammaOMO}{1-e^{-\gammaOMO\tau}}} \int_0^{\tau}ds\;e^{\frac{-\gammaOMO s}{2}} \ain(s), \\ 
\Aout = \sqrt{\frac{\gammaOMO}{e^{\gammaOMO\tau}-1}} \int_0^{\tau}ds\;e^{\frac{\gammaOMO s}{2}} \aout(s) \\
\Fhat = \sqrt{\gammabO} e^{\frac{\gammaOMO \tau}{2}} \int_0^{\tau}ds\;e^{\frac{-\gammaOMO s}{2}} \bin(s),
\end{gather}
\noindent which allows us to rewrite the input/output equations at the end of the pulse ($t = \tau$)
\begin{gather}
\Aout = e^{\frac{\gammaOMO \tau}{2}} \Ain +i\sqrt{e^{\gammaOMO \tau}-1}\Bindag, \\
\Bout = e^{\frac{\gammaOMO \tau}{2}} \Bin +i\sqrt{e^{\gammaOMO \tau}-1}\Aindag +\Fhat.
\end{gather}

Note that $[\Ain,\Aindag] = [\Aout,\Aoutdag] = [\Bin,\Bindag] =1$ as expected. As for the noise operator $\Fhat$, the commutator and correlation functions are defined, using the known properties of $\bin(t)$ to lowest order as
\begin{gather}
[\Fhat,\Fdag] \approx \gammabO \tau +O(\gammaOMO\gammabO\tau^2), \\
\Fcorr = \gammabO e^{\gammaOMO \tau} \int_0^{\tau}ds \;  e^{-\gammaOMO s} \nbath(s), \quad \left<\Fhat\Fdag\right> = \gammabO \tau + \left<\Fdag\Fhat\right>.
\end{gather}

We now seek a unitary propagator $\Uhat$ such that $\Aout = \Udag \Ain \Uhat$ and $\Bout = \Udag \Bin \Uhat$. Note that if we define $e^{\gammaOMO \tau /2} = \text{cosh}(r), \sqrt{e^{\gammaOMO \tau} -1} = \text{sinh}(r)$, then the input/output equations  have the form of two-mode squeezing between $\Ain$ and $\Bin$ in the absence of mechanical noise $\Fhat$. Thus, we want $\Uhat = \Uhatm \UhatO$ where
\begin{gather}
\UhatO = e^{i r \left(\Ain\Bin+\Aindag\Bindag\right)}, \\
\Udagm \Bin \Uhatm = \Bin+\Fhat, \\
[\Uhatm,\Ain] = 0.
\end{gather}

We find that a beam-splitter type interaction between $\Bin$ and $\Fhat$ satisfies this condition to lowest order. That is, if $\Uhatm = e^{\Bindag \Fhat-\Bin\Fdag}$, then using the above commutation relations for $\Fhat$ we can show,
\begin{align}
\Udagm\Bin\Uhatm & = \text{cos}\left(\sqrt{\gammabO\tau}\right)\Bin+\left(\gammabO\tau\right)^{-1/2}\text{sin}\left(\sqrt{\gammabO\tau}\right)\Fhat \nonumber\\
& \approx \Bin + \Fhat.
\end{align}

Given this propagator $\Uhat$ which yields the approximate system evolution over the pulse interval $\tau$, and assuming an initial density matrix $\rho_0 = \left|0_\text{A}\right>\left<0_\text{A}\right|\otimes\rho_\text{B,0}\otimes\rho_\text{F}$, where A, B, and F refer to the optical, mechanical, and bath subsystems respectively, the density matrix for the mechanical resonator, $\rho_\text{B}$, conditioned upon detection of a single photon, is given by 
\begin{align}
\rho_\text{B} & = \frac{\text{Tr}_\text{A,F} \left(\left|1_\text{A}\right>\left<1_\text{A}\right|\otimes I_\text{B}\otimes I_\text{F}\Uhat\rho_0\Udag\right)}{\text{Tr}\left(\left|1_\text{A}\right>\left<1_\text{A}\right|\otimes I_\text{B}\otimes I_\text{F}\rho_0\right)} \nonumber \\
& = P^{-1} \text{Tr}_\text{F} \left(\Uhatm \left<1_\text{A}\right|\UhatO\rho_0\UdagO\left|1_\text{A}\right>\Udagm\right),
\end{align}

\noindent where $P$ is the overall probability of photodetection, given by
\begin{equation}
P = \text{Tr}_\text{B}\left(\left<1_\text{A}\right|\UhatO\left|0_\text{A}\right>\left<0_\text{A}\right|\otimes\rho_\text{B,0}\UdagO\left|1_\text{A}\right>\right).
\end{equation}

Denoting $\rho_\text{B,ideal}  = P^{-1}\Bindag\rho_\text{B,0}\Bin$ as the ideal conditional matrix in the absence of mechanical dissipation, and expanding $\Uhat$ to lowest order in $\gammaOMO\tau$, $\gammabO\tau$ and $\Fcorr$, we find
\begin{equation}
P \approx \gammaOMO\tau \left<\Bindag\Bin\right>,
\end{equation}
\noindent and
\begin{align}
\rho_\text{B} & \approx \left(1-\Fcorr\right)\rho_\text{B,ideal}+\Fcorr \Bindag\rho_\text{B,ideal}\Bin +\left(\gammabO\tau+\Fcorr\right)\Bin\rho_\text{B,ideal}\Bindag \nonumber \\
& \quad -\frac{1}{2}\left(\gammabO\tau+2\Fcorr\right)\left(\Bindag\Bin\rho_\text{B,ideal}+\rho_\text{B,ideal}\Bindag\Bin\right).
\label{eqn:rho_B}
\end{align}

To evaluate the fidelity of the phonon addition subtraction, we use the definition of fidelity between two quantum states $\rho_1$ and $\rho_2$, given by~\cite{Nielsen2000} $F = \text{Tr}\left(\sqrt{\rho_1^{1/2}\rho_2\rho_1^{1/2}}\right)$, which is straightforward to evaluate for an initial thermal mechanical state, as both $\rho_1$ and $\rho_2$ will be diagonal in the Fock state basis.

Note that the only substantive difference in the case of phonon subtraction ($Delta = \omegamO$, red-detuned pumping) is that the propagator $\UhatO$ will have the form of a beam-splitter interaction rather than two mode squeezing. This will only change $P$ and $\rho_\text{B,ideal}$, which are now given by
\begin{gather}
P \approx \gammaOMO\tau\left<\Bindag\Bin\right>, \\
\rho_\text{B,ideal} = \Bin\rho_\text{B,0}\Bindag,
\end{gather}
\noindent while the definition of $\rho_\text{B}$ in terms of $\rho_\text{B,ideal}$ is unchanged assuming that the bath heating is only dependent on pump power and not pump detuning.

\begin{figure}[btp]
\begin{center}
\includegraphics[width=\textwidth]{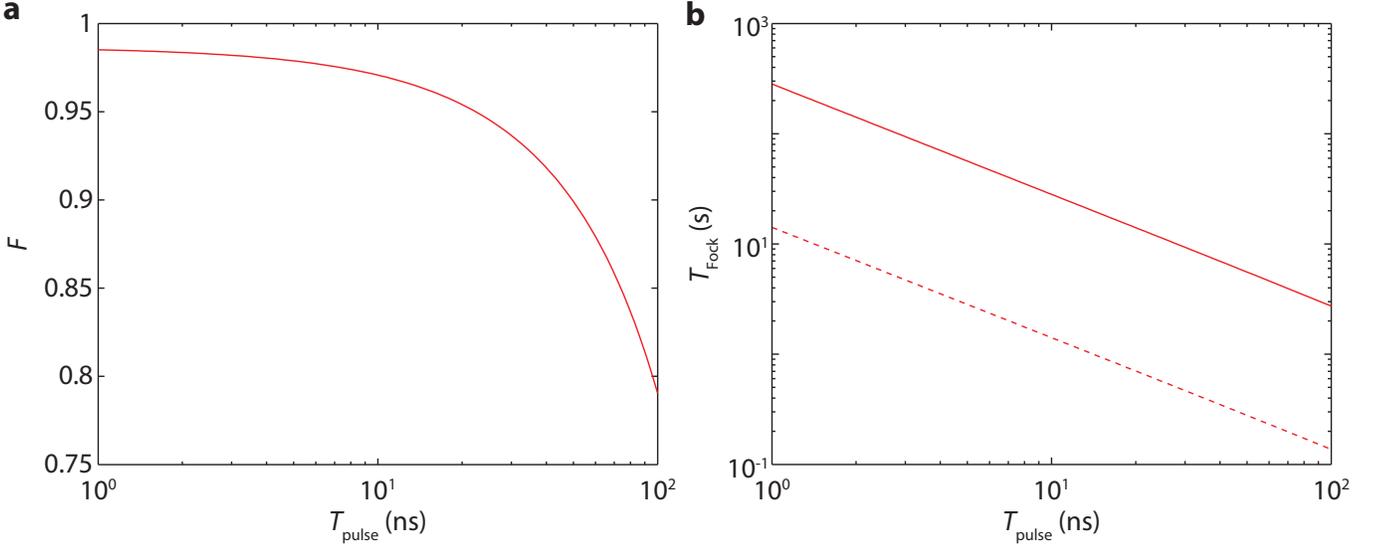}
\caption{\textbf{Fock state generation.} \textbf{a}, Fidelity $F$ of the generation of a single phonon Fock state versus pulse width $\Tpulse$ (pulse period $\Tper = 1$~ms). \textbf{b}, Average time required to herald a Fock state $T_\text{Fock}$ versus pulse width $\Tpulse$. The solid line is calculated using the total detection efficiency measured in this work, while the dashed line uses the estimated ideal detection efficiency.} \label{fig:fidelity_SI}
\end{center} 
\end{figure}

To evaluate the fidelity of the creation of a single phonon Fock state~\cite{Galland2014} in our system as a function of pulse width and period for blue-detuned pulses, we assume the initial state of the mechanics to be a thermal state with average phonon number $\nbar$ given by Eq.~\ref{eqn:n_init}, while $\Fcorr$ can be evaluated using the fit parameters of the effective time-dependent hot phonon bath. We may then calculate the conditional density matrix via Eq.~\ref{eqn:rho_B} and the fidelity by setting $\rho_1 = \left|1_\text{B}\right>\left<1_\text{B}\right|$ and $\rho_2 = \rho_\text{B}$. Considering a range of pulse widths $0 < \Tpulse \leq 100$~ns, which is sufficient to guarantee $\gammaOMO \Tpulse \ll 1$, we find that for $\Tper \geq 1$~ms there is no appreciable loss in fidelity due to long term heating causing an increase in $\nbar (t_0)$, and the primary loss in fidelity is due to transient heating during the pulse. Thus, we set the $\Tper = 1$~ms  and calculate the fidelity of Fock state generation as a function of pulse width, displayed in Fig.~\ref{fig:fidelity_SI}a. For short ($\Tpulse < 10$~ns) pulse widths, the fidelity approaches $98.5\%$, and remains above $80\%$ for pulse widths approaching 100 ns. However, it is equally important to quantify the expected time to herald such a state (i.e. the expected time before detection of a sideband photon), which in this case is given by $T_\text{Fock} = \Tper/(\eta \gammaOMO \Tpulse \nbar (t_0))$, where $\eta$ is the total detection efficiency of the phonon counting measurement. This is shown versus $\Tpulse$ in Fig.~\ref{fig:fidelity_SI}b for two cases. The solid line is calculated the actual measured detection efficiency of $\eta \approx 0.3\%$, while the dashed line is calculated using a realistic estimate of the ideal efficiency of our current measurement setup, $\eta \approx 5.5\%$. This latter figure is calculated using the measured efficiencies of the SPD ($\etaSPD = 68\%$), the fiber-to-waveguide coupling efficiency ($\eta_\text{cpl} = 68\%$), and the waveguide-cavity coupling efficiency ($\eta_\kappa = 50\%$), and adding an additional 2 dB insertion loss per filter and 0.5 dB for the insertion loss of the optical circulator, which correspond to the highest efficiencies measured for these components in our lab. As shown in Fig.~\ref{fig:fidelity_SI}, even in the idealized case the expected time for the generation of a Fock state is $\gtrsim 100$~ms, which is much longer than the lifetime of the mechanical state. Thus, while it is feasible to use our current systems for the heralded, high-fidelity generation of non-classical mechanical states, it it still necessary to reduce heating effects (and thus allow for shorter pulse periods) in order to utilize this procedure for useful quantum information processing tasks (e.g. scalable entanglement distribution via the DLCZ protocol~\cite{Duan2001,Sangouard2011}).


\begin{thebibliography}{37}%
\makeatletter
\providecommand \@ifxundefined [1]{%
 \@ifx{#1\undefined}
}%
\providecommand \@ifnum [1]{%
 \ifnum #1\expandafter \@firstoftwo
 \else \expandafter \@secondoftwo
 \fi
}%
\providecommand \@ifx [1]{%
 \ifx #1\expandafter \@firstoftwo
 \else \expandafter \@secondoftwo
 \fi
}%
\providecommand \natexlab [1]{#1}%
\providecommand \enquote  [1]{``#1''}%
\providecommand \bibnamefont  [1]{#1}%
\providecommand \bibfnamefont [1]{#1}%
\providecommand \citenamefont [1]{#1}%
\providecommand \href@noop [0]{\@secondoftwo}%
\providecommand \href [0]{\begingroup \@sanitize@url \@href}%
\providecommand \@href[1]{\@@startlink{#1}\@@href}%
\providecommand \@@href[1]{\endgroup#1\@@endlink}%
\providecommand \@sanitize@url [0]{\catcode `\\12\catcode `\$12\catcode
  `\&12\catcode `\#12\catcode `\^12\catcode `\_12\catcode `\%12\relax}%
\providecommand \@@startlink[1]{}%
\providecommand \@@endlink[0]{}%
\providecommand \url  [0]{\begingroup\@sanitize@url \@url }%
\providecommand \@url [1]{\endgroup\@href {#1}{\urlprefix }}%
\providecommand \urlprefix  [0]{URL }%
\providecommand \Eprint [0]{\href }%
\providecommand \doibase [0]{http://dx.doi.org/}%
\providecommand \selectlanguage [0]{\@gobble}%
\providecommand \bibinfo  [0]{\@secondoftwo}%
\providecommand \bibfield  [0]{\@secondoftwo}%
\providecommand \translation [1]{[#1]}%
\providecommand \BibitemOpen [0]{}%
\providecommand \bibitemStop [0]{}%
\providecommand \bibitemNoStop [0]{.\EOS\space}%
\providecommand \EOS [0]{\spacefactor3000\relax}%
\providecommand \BibitemShut  [1]{\csname bibitem#1\endcsname}%
\let\auto@bib@innerbib\@empty
%</preamble>
\bibitem [{\citenamefont {O'Connell}\ \emph {et~al.}(2010)\citenamefont
  {O'Connell}, \citenamefont {Hofheinz}, \citenamefont {Ansmann}, \citenamefont
  {Bialczak}, \citenamefont {Lenander}, \citenamefont {Lucero}, \citenamefont
  {Neeley}, \citenamefont {Sank}, \citenamefont {Wang}, \citenamefont {Weides},
  \citenamefont {Wenner}, \citenamefont {Martinis},\ and\ \citenamefont
  {Cleland}}]{OConnell2010}%
  \BibitemOpen
  \bibfield  {author} {\bibinfo {author} {\bibfnamefont {A.~D.}\ \bibnamefont
  {O'Connell}}, \bibinfo {author} {\bibfnamefont {M.}~\bibnamefont {Hofheinz}},
  \bibinfo {author} {\bibfnamefont {M.}~\bibnamefont {Ansmann}}, \bibinfo
  {author} {\bibfnamefont {R.~C.}\ \bibnamefont {Bialczak}}, \bibinfo {author}
  {\bibfnamefont {M.}~\bibnamefont {Lenander}}, \bibinfo {author}
  {\bibfnamefont {E.}~\bibnamefont {Lucero}}, \bibinfo {author} {\bibfnamefont
  {M.}~\bibnamefont {Neeley}}, \bibinfo {author} {\bibfnamefont
  {D.}~\bibnamefont {Sank}}, \bibinfo {author} {\bibfnamefont {H.}~\bibnamefont
  {Wang}}, \bibinfo {author} {\bibfnamefont {M.}~\bibnamefont {Weides}},
  \bibinfo {author} {\bibfnamefont {J.}~\bibnamefont {Wenner}}, \bibinfo
  {author} {\bibfnamefont {J.~M.}\ \bibnamefont {Martinis}}, \ and\ \bibinfo
  {author} {\bibfnamefont {A.~N.}\ \bibnamefont {Cleland}},\ }\href@noop {}
  {\bibfield  {journal} {\bibinfo  {journal} {Nature}\ }\textbf {\bibinfo
  {volume} {464}},\ \bibinfo {pages} {697} (\bibinfo {year}
  {2010})}\BibitemShut {NoStop}%
\bibitem [{\citenamefont {Teufel}\ \emph {et~al.}(2011)\citenamefont {Teufel},
  \citenamefont {Donner}, \citenamefont {Li}, \citenamefont {Harlow},
  \citenamefont {Allman}, \citenamefont {Cicak}, \citenamefont {Sirois},
  \citenamefont {Whittaker}, \citenamefont {Lehnert},\ and\ \citenamefont
  {Simmonds}}]{Teufel2011b}%
  \BibitemOpen
  \bibfield  {author} {\bibinfo {author} {\bibfnamefont {J.~D.}\ \bibnamefont
  {Teufel}}, \bibinfo {author} {\bibfnamefont {T.}~\bibnamefont {Donner}},
  \bibinfo {author} {\bibfnamefont {D.}~\bibnamefont {Li}}, \bibinfo {author}
  {\bibfnamefont {J.~W.}\ \bibnamefont {Harlow}}, \bibinfo {author}
  {\bibfnamefont {M.~S.}\ \bibnamefont {Allman}}, \bibinfo {author}
  {\bibfnamefont {K.}~\bibnamefont {Cicak}}, \bibinfo {author} {\bibfnamefont
  {A.~J.}\ \bibnamefont {Sirois}}, \bibinfo {author} {\bibfnamefont {J.~D.}\
  \bibnamefont {Whittaker}}, \bibinfo {author} {\bibfnamefont {K.~W.}\
  \bibnamefont {Lehnert}}, \ and\ \bibinfo {author} {\bibfnamefont {R.~W.}\
  \bibnamefont {Simmonds}},\ }\href@noop {} {\bibfield  {journal} {\bibinfo
  {journal} {Nature}\ }\textbf {\bibinfo {volume} {475}},\ \bibinfo {pages}
  {359} (\bibinfo {year} {2011})}\BibitemShut {NoStop}%
\bibitem [{\citenamefont {Chan}\ \emph {et~al.}(2011)\citenamefont {Chan},
  \citenamefont {Alegre}, \citenamefont {Safavi-Naeini}, \citenamefont {Hill},
  \citenamefont {Krause}, \citenamefont {Gr\"oblacher}, \citenamefont
  {Aspelmeyer},\ and\ \citenamefont {Painter}}]{Chan2011}%
  \BibitemOpen
  \bibfield  {author} {\bibinfo {author} {\bibfnamefont {J.}~\bibnamefont
  {Chan}}, \bibinfo {author} {\bibfnamefont {T.~P.~M.}\ \bibnamefont {Alegre}},
  \bibinfo {author} {\bibfnamefont {A.~H.}\ \bibnamefont {Safavi-Naeini}},
  \bibinfo {author} {\bibfnamefont {J.~T.}\ \bibnamefont {Hill}}, \bibinfo
  {author} {\bibfnamefont {A.}~\bibnamefont {Krause}}, \bibinfo {author}
  {\bibfnamefont {S.}~\bibnamefont {Gr\"oblacher}}, \bibinfo {author}
  {\bibfnamefont {M.}~\bibnamefont {Aspelmeyer}}, \ and\ \bibinfo {author}
  {\bibfnamefont {O.}~\bibnamefont {Painter}},\ }\href@noop {} {\bibfield
  {journal} {\bibinfo  {journal} {Nature}\ }\textbf {\bibinfo {volume} {478}},\
  \bibinfo {pages} {89} (\bibinfo {year} {2011})}\BibitemShut {NoStop}%
\bibitem [{\citenamefont {Aspelmeyer}\ \emph {et~al.}(2014)\citenamefont
  {Aspelmeyer}, \citenamefont {Kippenberg},\ and\ \citenamefont
  {Marquardt}}]{Aspelmeyer2014}%
  \BibitemOpen
  \bibfield  {author} {\bibinfo {author} {\bibfnamefont {M.}~\bibnamefont
  {Aspelmeyer}}, \bibinfo {author} {\bibfnamefont {T.~J.}\ \bibnamefont
  {Kippenberg}}, \ and\ \bibinfo {author} {\bibfnamefont {F.}~\bibnamefont
  {Marquardt}},\ }\href@noop {} {\bibfield  {journal} {\bibinfo  {journal}
  {Rev. Mod. Phys.}\ }\textbf {\bibinfo {volume} {86}},\ \bibinfo {pages}
  {1391} (\bibinfo {year} {2014})}\BibitemShut {NoStop}%
\bibitem [{\citenamefont {Vanner}\ \emph {et~al.}(2013)\citenamefont {Vanner},
  \citenamefont {Aspelmeyer},\ and\ \citenamefont {Kim}}]{Vanner2013}%
  \BibitemOpen
  \bibfield  {author} {\bibinfo {author} {\bibfnamefont {M.~R.}\ \bibnamefont
  {Vanner}}, \bibinfo {author} {\bibfnamefont {M.}~\bibnamefont {Aspelmeyer}},
  \ and\ \bibinfo {author} {\bibfnamefont {M.~S.}\ \bibnamefont {Kim}},\
  }\href@noop {} {\bibfield  {journal} {\bibinfo  {journal} {Phys.\ Rev.\
  Lett.}\ }\textbf {\bibinfo {volume} {110}},\ \bibinfo {pages} {010504}
  (\bibinfo {year} {2013})}\BibitemShut {NoStop}%
\bibitem [{\citenamefont {B{\o}rkje}\ \emph {et~al.}(2011)\citenamefont
  {B{\o}rkje}, \citenamefont {Nunnenkamp},\ and\ \citenamefont
  {Girvin}}]{Borkje2011}%
  \BibitemOpen
  \bibfield  {author} {\bibinfo {author} {\bibfnamefont {K.}~\bibnamefont
  {B{\o}rkje}}, \bibinfo {author} {\bibfnamefont {A.}~\bibnamefont
  {Nunnenkamp}}, \ and\ \bibinfo {author} {\bibfnamefont {S.~M.}\ \bibnamefont
  {Girvin}},\ }\href@noop {} {\bibfield  {journal} {\bibinfo  {journal} {Phys.\
  Rev.\ Lett.}\ }\textbf {\bibinfo {volume} {107}},\ \bibinfo {pages} {123601}
  (\bibinfo {year} {2011})}\BibitemShut {NoStop}%
\bibitem [{\citenamefont {Galland}\ \emph {et~al.}(2014)\citenamefont
  {Galland}, \citenamefont {Sangouard}, \citenamefont {Piro}, \citenamefont
  {Gisin},\ and\ \citenamefont {Kippenberg}}]{Galland2014}%
  \BibitemOpen
  \bibfield  {author} {\bibinfo {author} {\bibfnamefont {C.}~\bibnamefont
  {Galland}}, \bibinfo {author} {\bibfnamefont {N.}~\bibnamefont {Sangouard}},
  \bibinfo {author} {\bibfnamefont {N.}~\bibnamefont {Piro}}, \bibinfo {author}
  {\bibfnamefont {N.}~\bibnamefont {Gisin}}, \ and\ \bibinfo {author}
  {\bibfnamefont {T.~J.}\ \bibnamefont {Kippenberg}},\ }\href@noop {}
  {\bibfield  {journal} {\bibinfo  {journal} {Phys. Rev. Lett.}\ }\textbf
  {\bibinfo {volume} {112}},\ \bibinfo {pages} {143602} (\bibinfo {year}
  {2014})}\BibitemShut {NoStop}%
\bibitem [{\citenamefont {Stannigel}\ \emph {et~al.}(2010)\citenamefont
  {Stannigel}, \citenamefont {Rabl}, \citenamefont {S\o{}rensen}, \citenamefont
  {Zoller},\ and\ \citenamefont {Lukin}}]{Stannigel2010}%
  \BibitemOpen
  \bibfield  {author} {\bibinfo {author} {\bibfnamefont {K.}~\bibnamefont
  {Stannigel}}, \bibinfo {author} {\bibfnamefont {P.}~\bibnamefont {Rabl}},
  \bibinfo {author} {\bibfnamefont {A.~S.}\ \bibnamefont {S\o{}rensen}},
  \bibinfo {author} {\bibfnamefont {P.}~\bibnamefont {Zoller}}, \ and\ \bibinfo
  {author} {\bibfnamefont {M.~D.}\ \bibnamefont {Lukin}},\ }\href@noop {}
  {\bibfield  {journal} {\bibinfo  {journal} {Phys.\ Rev.\ Lett.}\ }\textbf
  {\bibinfo {volume} {105}},\ \bibinfo {pages} {220501} (\bibinfo {year}
  {2010})}\BibitemShut {NoStop}%
\bibitem [{\citenamefont {Safavi-Naeini}\ and\ \citenamefont
  {Painter}(2011)}]{Safavi-Naeini2011a}%
  \BibitemOpen
  \bibfield  {author} {\bibinfo {author} {\bibfnamefont {A.~H.}\ \bibnamefont
  {Safavi-Naeini}}\ and\ \bibinfo {author} {\bibfnamefont {O.}~\bibnamefont
  {Painter}},\ }\href@noop {} {\bibfield  {journal} {\bibinfo  {journal} {New
  J.\ Phys.}\ }\textbf {\bibinfo {volume} {13}},\ \bibinfo {pages} {013017}
  (\bibinfo {year} {2011})}\BibitemShut {NoStop}%
\bibitem [{\citenamefont {Hill}\ \emph {et~al.}(2012)\citenamefont {Hill},
  \citenamefont {Safavi-Naeini}, \citenamefont {Chan},\ and\ \citenamefont
  {Painter}}]{Hill2012}%
  \BibitemOpen
  \bibfield  {author} {\bibinfo {author} {\bibfnamefont {J.~T.}\ \bibnamefont
  {Hill}}, \bibinfo {author} {\bibfnamefont {A.~H.}\ \bibnamefont
  {Safavi-Naeini}}, \bibinfo {author} {\bibfnamefont {J.}~\bibnamefont {Chan}},
  \ and\ \bibinfo {author} {\bibfnamefont {O.}~\bibnamefont {Painter}},\
  }\href@noop {} {\bibfield  {journal} {\bibinfo  {journal} {Nature Commun.}\
  }\textbf {\bibinfo {volume} {3}},\ \bibinfo {pages} {1196} (\bibinfo {year}
  {2012})}\BibitemShut {NoStop}%
\bibitem [{\citenamefont {Bochmann}\ \emph {et~al.}(2013)\citenamefont
  {Bochmann}, \citenamefont {Vainsencher}, \citenamefont {Awschalom},\ and\
  \citenamefont {Cleland}}]{Bochmann2013}%
  \BibitemOpen
  \bibfield  {author} {\bibinfo {author} {\bibfnamefont {J.}~\bibnamefont
  {Bochmann}}, \bibinfo {author} {\bibfnamefont {A.}~\bibnamefont
  {Vainsencher}}, \bibinfo {author} {\bibfnamefont {D.~D.}\ \bibnamefont
  {Awschalom}}, \ and\ \bibinfo {author} {\bibfnamefont {A.~N.}\ \bibnamefont
  {Cleland}},\ }\href@noop {} {\bibfield  {journal} {\bibinfo  {journal}
  {Nature Physics}\ }\textbf {\bibinfo {volume} {9}},\ \bibinfo {pages} {712}
  (\bibinfo {year} {2013})}\BibitemShut {NoStop}%
\bibitem [{\citenamefont {Andrews}\ \emph {et~al.}(2014)\citenamefont
  {Andrews}, \citenamefont {Peterson}, \citenamefont {Purdy}, \citenamefont
  {Cicak}, \citenamefont {Simmonds}, \citenamefont {Regal},\ and\ \citenamefont
  {Lehnert}}]{Andrews2014}%
  \BibitemOpen
  \bibfield  {author} {\bibinfo {author} {\bibfnamefont {R.~W.}\ \bibnamefont
  {Andrews}}, \bibinfo {author} {\bibfnamefont {R.~W.}\ \bibnamefont
  {Peterson}}, \bibinfo {author} {\bibfnamefont {T.~P.}\ \bibnamefont {Purdy}},
  \bibinfo {author} {\bibfnamefont {K.}~\bibnamefont {Cicak}}, \bibinfo
  {author} {\bibfnamefont {R.~W.}\ \bibnamefont {Simmonds}}, \bibinfo {author}
  {\bibfnamefont {C.~A.}\ \bibnamefont {Regal}}, \ and\ \bibinfo {author}
  {\bibfnamefont {K.}~\bibnamefont {Lehnert}},\ }\href@noop {} {\bibfield
  {journal} {\bibinfo  {journal} {Nature Physics}\ }\textbf {\bibinfo {volume}
  {10}},\ \bibinfo {pages} {321} (\bibinfo {year} {2014})}\BibitemShut
  {NoStop}%
\bibitem [{\citenamefont {Eichenfield}\ \emph {et~al.}(2009)\citenamefont
  {Eichenfield}, \citenamefont {Chan}, \citenamefont {Camacho}, \citenamefont
  {Vahala},\ and\ \citenamefont {Painter}}]{Eichenfield2009b}%
  \BibitemOpen
  \bibfield  {author} {\bibinfo {author} {\bibfnamefont {M.}~\bibnamefont
  {Eichenfield}}, \bibinfo {author} {\bibfnamefont {J.}~\bibnamefont {Chan}},
  \bibinfo {author} {\bibfnamefont {R.~M.}\ \bibnamefont {Camacho}}, \bibinfo
  {author} {\bibfnamefont {K.~J.}\ \bibnamefont {Vahala}}, \ and\ \bibinfo
  {author} {\bibfnamefont {O.}~\bibnamefont {Painter}},\ }\href@noop {}
  {\bibfield  {journal} {\bibinfo  {journal} {Nature}\ }\textbf {\bibinfo
  {volume} {462}},\ \bibinfo {pages} {78} (\bibinfo {year} {2009})}\BibitemShut
  {NoStop}%
\bibitem [{\citenamefont {Safavi-Naeini}\ and\ \citenamefont
  {Painter}(2010)}]{Safavi-Naeini2010b}%
  \BibitemOpen
  \bibfield  {author} {\bibinfo {author} {\bibfnamefont {A.~H.}\ \bibnamefont
  {Safavi-Naeini}}\ and\ \bibinfo {author} {\bibfnamefont {O.}~\bibnamefont
  {Painter}},\ }\href@noop {} {\bibfield  {journal} {\bibinfo  {journal} {Opt.
  Express}\ }\textbf {\bibinfo {volume} {18}},\ \bibinfo {pages} {14926}
  (\bibinfo {year} {2010})}\BibitemShut {NoStop}%
\bibitem [{\citenamefont {Habraken}\ \emph {et~al.}(2012)\citenamefont
  {Habraken}, \citenamefont {Stannigel}, \citenamefont {Lukin}, \citenamefont
  {Zoller},\ and\ \citenamefont {Rabl}}]{Habraken2012}%
  \BibitemOpen
  \bibfield  {author} {\bibinfo {author} {\bibfnamefont {S.~J.~M.}\
  \bibnamefont {Habraken}}, \bibinfo {author} {\bibfnamefont {K.}~\bibnamefont
  {Stannigel}}, \bibinfo {author} {\bibfnamefont {M.~D.}\ \bibnamefont
  {Lukin}}, \bibinfo {author} {\bibfnamefont {P.}~\bibnamefont {Zoller}}, \
  and\ \bibinfo {author} {\bibfnamefont {P.}~\bibnamefont {Rabl}},\ }\href@noop
  {} {\bibfield  {journal} {\bibinfo  {journal} {New J.\ Phys.}\ ,\ \bibinfo
  {pages} {115004}} (\bibinfo {year} {2012})}\BibitemShut {NoStop}%
\bibitem [{\citenamefont {Schmidt}\ \emph {et~al.}(2013)\citenamefont
  {Schmidt}, \citenamefont {Peano},\ and\ \citenamefont
  {Marquardt}}]{Schmidt2013}%
  \BibitemOpen
  \bibfield  {author} {\bibinfo {author} {\bibfnamefont {M.}~\bibnamefont
  {Schmidt}}, \bibinfo {author} {\bibfnamefont {V.}~\bibnamefont {Peano}}, \
  and\ \bibinfo {author} {\bibfnamefont {F.}~\bibnamefont {Marquardt}},\
  }\href@noop {} {\bibfield  {journal} {\bibinfo  {journal} {arXiv:1311.7095}\
  } (\bibinfo {year} {2013})}\BibitemShut {NoStop}%
\bibitem [{\citenamefont {Chan}\ \emph {et~al.}(2012)\citenamefont {Chan},
  \citenamefont {Safavi-Naeini}, \citenamefont {Hill}, \citenamefont
  {Meenehan},\ and\ \citenamefont {Painter}}]{Chan2012}%
  \BibitemOpen
  \bibfield  {author} {\bibinfo {author} {\bibfnamefont {J.}~\bibnamefont
  {Chan}}, \bibinfo {author} {\bibfnamefont {A.~H.}\ \bibnamefont
  {Safavi-Naeini}}, \bibinfo {author} {\bibfnamefont {J.~T.}\ \bibnamefont
  {Hill}}, \bibinfo {author} {\bibfnamefont {S.}~\bibnamefont {Meenehan}}, \
  and\ \bibinfo {author} {\bibfnamefont {O.}~\bibnamefont {Painter}},\
  }\href@noop {} {\bibfield  {journal} {\bibinfo  {journal} {Appl.\ Phys.\
  Lett.}\ }\textbf {\bibinfo {volume} {101}},\ \bibinfo {pages} {081115}
  (\bibinfo {year} {2012})}\BibitemShut {NoStop}%
\bibitem [{\citenamefont {Meenehan}\ \emph {et~al.}(2014)\citenamefont
  {Meenehan}, \citenamefont {Cohen}, \citenamefont {Gr\"{o}blacher},
  \citenamefont {Hill}, \citenamefont {Safavi-Naeini}, \citenamefont
  {Aspelmeyer},\ and\ \citenamefont {Painter}}]{Meenehan2014}%
  \BibitemOpen
  \bibfield  {author} {\bibinfo {author} {\bibfnamefont {S.~M.}\ \bibnamefont
  {Meenehan}}, \bibinfo {author} {\bibfnamefont {J.~D.}\ \bibnamefont {Cohen}},
  \bibinfo {author} {\bibfnamefont {S.}~\bibnamefont {Gr\"{o}blacher}},
  \bibinfo {author} {\bibfnamefont {J.~T.}\ \bibnamefont {Hill}}, \bibinfo
  {author} {\bibfnamefont {A.~H.}\ \bibnamefont {Safavi-Naeini}}, \bibinfo
  {author} {\bibfnamefont {M.}~\bibnamefont {Aspelmeyer}}, \ and\ \bibinfo
  {author} {\bibfnamefont {O.}~\bibnamefont {Painter}},\ }\href@noop {}
  {\bibfield  {journal} {\bibinfo  {journal} {Phys. Rev. A}\ }\textbf {\bibinfo
  {volume} {90}},\ \bibinfo {pages} {011803} (\bibinfo {year}
  {2014})}\BibitemShut {NoStop}%
\bibitem [{\citenamefont {Cohen}\ \emph {et~al.}(2014)\citenamefont {Cohen},
  \citenamefont {Meenehan}, \citenamefont {MacCabe}, \citenamefont
  {Gr\"oblacher}, \citenamefont {Safavi-Naeini}, \citenamefont {Marsili},
  \citenamefont {Shaw},\ and\ \citenamefont {Painter}}]{Cohen2014}%
  \BibitemOpen
  \bibfield  {author} {\bibinfo {author} {\bibfnamefont {J.~D.}\ \bibnamefont
  {Cohen}}, \bibinfo {author} {\bibfnamefont {S.~M.}\ \bibnamefont {Meenehan}},
  \bibinfo {author} {\bibfnamefont {G.~S.}\ \bibnamefont {MacCabe}}, \bibinfo
  {author} {\bibfnamefont {S.}~\bibnamefont {Gr\"oblacher}}, \bibinfo {author}
  {\bibfnamefont {A.~H.}\ \bibnamefont {Safavi-Naeini}}, \bibinfo {author}
  {\bibfnamefont {F.}~\bibnamefont {Marsili}}, \bibinfo {author} {\bibfnamefont
  {M.~D.}\ \bibnamefont {Shaw}}, \ and\ \bibinfo {author} {\bibfnamefont
  {O.}~\bibnamefont {Painter}},\ }\href@noop {} {\bibfield  {journal} {\bibinfo
   {journal} {arXiv:1410.1047}\ } (\bibinfo {year} {2014})}\BibitemShut
  {NoStop}%
\bibitem [{\citenamefont {Safavi-Naeini}\ \emph {et~al.}(2012)\citenamefont
  {Safavi-Naeini}, \citenamefont {Chan}, \citenamefont {Hill}, \citenamefont
  {Alegre}, \citenamefont {Krause},\ and\ \citenamefont
  {Painter}}]{Safavi-Naeini2012}%
  \BibitemOpen
  \bibfield  {author} {\bibinfo {author} {\bibfnamefont {A.~H.}\ \bibnamefont
  {Safavi-Naeini}}, \bibinfo {author} {\bibfnamefont {J.}~\bibnamefont {Chan}},
  \bibinfo {author} {\bibfnamefont {J.~T.}\ \bibnamefont {Hill}}, \bibinfo
  {author} {\bibfnamefont {T.~P.~M.}\ \bibnamefont {Alegre}}, \bibinfo {author}
  {\bibfnamefont {A.}~\bibnamefont {Krause}}, \ and\ \bibinfo {author}
  {\bibfnamefont {O.}~\bibnamefont {Painter}},\ }\href@noop {} {\bibfield
  {journal} {\bibinfo  {journal} {Phys. Rev. Lett.}\ }\textbf {\bibinfo
  {volume} {108}},\ \bibinfo {pages} {033602} (\bibinfo {year}
  {2012})}\BibitemShut {NoStop}%
\bibitem [{\citenamefont {Purdy}\ \emph {et~al.}(2014)\citenamefont {Purdy},
  \citenamefont {Yu}, \citenamefont {Kampel}, \citenamefont {Peterson},
  \citenamefont {Cicak}, \citenamefont {Simmonds},\ and\ \citenamefont
  {Regal}}]{Purdy2014}%
  \BibitemOpen
  \bibfield  {author} {\bibinfo {author} {\bibfnamefont {T.~P.}\ \bibnamefont
  {Purdy}}, \bibinfo {author} {\bibfnamefont {P.-L.}\ \bibnamefont {Yu}},
  \bibinfo {author} {\bibfnamefont {N.~S.}\ \bibnamefont {Kampel}}, \bibinfo
  {author} {\bibfnamefont {R.~W.}\ \bibnamefont {Peterson}}, \bibinfo {author}
  {\bibfnamefont {K.}~\bibnamefont {Cicak}}, \bibinfo {author} {\bibfnamefont
  {R.~W.}\ \bibnamefont {Simmonds}}, \ and\ \bibinfo {author} {\bibfnamefont
  {C.~A.}\ \bibnamefont {Regal}},\ }\href@noop {} {\bibfield  {journal}
  {\bibinfo  {journal} {arXiv:1406.7247}\ ,\ \bibinfo {pages} {032325}}
  (\bibinfo {year} {2014})}\BibitemShut {NoStop}%
\bibitem [{\citenamefont {Weinstein}\ \emph {et~al.}(2014)\citenamefont
  {Weinstein}, \citenamefont {Lei}, \citenamefont {Wollman}, \citenamefont
  {Suh}, \citenamefont {Metelmann}, \citenamefont {Clerk},\ and\ \citenamefont
  {Schwab}}]{Weinstein2014}%
  \BibitemOpen
  \bibfield  {author} {\bibinfo {author} {\bibfnamefont {A.~J.}\ \bibnamefont
  {Weinstein}}, \bibinfo {author} {\bibfnamefont {C.~U.}\ \bibnamefont {Lei}},
  \bibinfo {author} {\bibfnamefont {E.~E.}\ \bibnamefont {Wollman}}, \bibinfo
  {author} {\bibfnamefont {J.}~\bibnamefont {Suh}}, \bibinfo {author}
  {\bibfnamefont {A.}~\bibnamefont {Metelmann}}, \bibinfo {author}
  {\bibfnamefont {A.~A.}\ \bibnamefont {Clerk}}, \ and\ \bibinfo {author}
  {\bibfnamefont {K.~C.}\ \bibnamefont {Schwab}},\ }\href@noop {} {\bibfield
  {journal} {\bibinfo  {journal} {Phys. Rev. X}\ }\textbf {\bibinfo {volume}
  {4}},\ \bibinfo {pages} {041003} (\bibinfo {year} {2014})}\BibitemShut
  {NoStop}%
\bibitem [{\citenamefont {Stesmans}(1996)}]{Stesmans1996}%
  \BibitemOpen
  \bibfield  {author} {\bibinfo {author} {\bibfnamefont {A.}~\bibnamefont
  {Stesmans}},\ }\href@noop {} {\bibfield  {journal} {\bibinfo  {journal}
  {App.\ Phys.\ Lett.}\ }\textbf {\bibinfo {volume} {68}} (\bibinfo {year}
  {1996})}\BibitemShut {NoStop}%
\bibitem [{\citenamefont {Borselli}\ \emph {et~al.}(2006)\citenamefont
  {Borselli}, \citenamefont {Johnson},\ and\ \citenamefont
  {Painter}}]{Borselli2006}%
  \BibitemOpen
  \bibfield  {author} {\bibinfo {author} {\bibfnamefont {M.}~\bibnamefont
  {Borselli}}, \bibinfo {author} {\bibfnamefont {T.~J.}\ \bibnamefont
  {Johnson}}, \ and\ \bibinfo {author} {\bibfnamefont {O.}~\bibnamefont
  {Painter}},\ }\href@noop {} {\bibfield  {journal} {\bibinfo  {journal} {App.\
  Phys.\ Lett.}\ }\textbf {\bibinfo {volume} {88}},\ \bibinfo {pages} {131114}
  (\bibinfo {year} {2006})}\BibitemShut {NoStop}%
\bibitem [{\citenamefont {Holland}(1963)}]{Holland1963}%
  \BibitemOpen
  \bibfield  {author} {\bibinfo {author} {\bibfnamefont {M.~G.}\ \bibnamefont
  {Holland}},\ }\href@noop {} {\bibfield  {journal} {\bibinfo  {journal} {Phys.
  Rev.}\ }\textbf {\bibinfo {volume} {132}},\ \bibinfo {pages} {2461} (\bibinfo
  {year} {1963})}\BibitemShut {NoStop}%
\bibitem [{\citenamefont {Lee}\ \emph {et~al.}(2014)\citenamefont {Lee},
  \citenamefont {Underwood}, \citenamefont {Mason}, \citenamefont {Shkarin},
  \citenamefont {B{\o}rkje}, \citenamefont {Girvin},\ and\ \citenamefont
  {Harris}}]{Lee2014}%
  \BibitemOpen
  \bibfield  {author} {\bibinfo {author} {\bibfnamefont {D.}~\bibnamefont
  {Lee}}, \bibinfo {author} {\bibfnamefont {M.}~\bibnamefont {Underwood}},
  \bibinfo {author} {\bibfnamefont {D.}~\bibnamefont {Mason}}, \bibinfo
  {author} {\bibfnamefont {A.~B.}\ \bibnamefont {Shkarin}}, \bibinfo {author}
  {\bibfnamefont {K.}~\bibnamefont {B{\o}rkje}}, \bibinfo {author}
  {\bibfnamefont {S.~M.}\ \bibnamefont {Girvin}}, \ and\ \bibinfo {author}
  {\bibfnamefont {J.~G.~E.}\ \bibnamefont {Harris}},\ }\href@noop {} {\bibfield
   {journal} {\bibinfo  {journal} {arXiv:1406.7254}\ } (\bibinfo {year}
  {2014})}\BibitemShut {NoStop}%
\bibitem [{\citenamefont {Khalili}\ \emph {et~al.}(2012)\citenamefont
  {Khalili}, \citenamefont {Miao}, \citenamefont {Yang}, \citenamefont
  {Safavi-Naeini}, \citenamefont {Painter},\ and\ \citenamefont
  {Chen}}]{Khalili2012}%
  \BibitemOpen
  \bibfield  {author} {\bibinfo {author} {\bibfnamefont {F.~Y.}\ \bibnamefont
  {Khalili}}, \bibinfo {author} {\bibfnamefont {H.}~\bibnamefont {Miao}},
  \bibinfo {author} {\bibfnamefont {H.}~\bibnamefont {Yang}}, \bibinfo {author}
  {\bibfnamefont {A.~H.}\ \bibnamefont {Safavi-Naeini}}, \bibinfo {author}
  {\bibfnamefont {O.}~\bibnamefont {Painter}}, \ and\ \bibinfo {author}
  {\bibfnamefont {Y.}~\bibnamefont {Chen}},\ }\href@noop {} {\bibfield
  {journal} {\bibinfo  {journal} {Phys.\ Rev.\ A}\ }\textbf {\bibinfo {volume}
  {86}},\ \bibinfo {pages} {033840} (\bibinfo {year} {2012})}\BibitemShut
  {NoStop}%
\bibitem [{\citenamefont {Kippenberg}\ and\ \citenamefont
  {Vahala}(2007)}]{Kippenberg2007}%
  \BibitemOpen
  \bibfield  {author} {\bibinfo {author} {\bibfnamefont {T.~J.}\ \bibnamefont
  {Kippenberg}}\ and\ \bibinfo {author} {\bibfnamefont {K.~J.}\ \bibnamefont
  {Vahala}},\ }\href@noop {} {\bibfield  {journal} {\bibinfo  {journal} {Opt.\
  Express}\ }\textbf {\bibinfo {volume} {15}},\ \bibinfo {pages} {17172}
  (\bibinfo {year} {2007})}\BibitemShut {NoStop}%
\bibitem [{\citenamefont {Diedrich}\ \emph {et~al.}(1989)\citenamefont
  {Diedrich}, \citenamefont {Bergquist}, \citenamefont {Itano},\ and\
  \citenamefont {Wineland}}]{Diedrich1989}%
  \BibitemOpen
  \bibfield  {author} {\bibinfo {author} {\bibfnamefont {F.}~\bibnamefont
  {Diedrich}}, \bibinfo {author} {\bibfnamefont {J.~C.}\ \bibnamefont
  {Bergquist}}, \bibinfo {author} {\bibfnamefont {W.~M.}\ \bibnamefont
  {Itano}}, \ and\ \bibinfo {author} {\bibfnamefont {D.~J.}\ \bibnamefont
  {Wineland}},\ }\href@noop {} {\bibfield  {journal} {\bibinfo  {journal}
  {Phys. Rev. Lett.}\ }\textbf {\bibinfo {volume} {62}},\ \bibinfo {pages}
  {403} (\bibinfo {year} {1989})}\BibitemShut {NoStop}%
\bibitem [{\citenamefont {Marsili}\ \emph {et~al.}(2013)\citenamefont
  {Marsili}, \citenamefont {Verma}, \citenamefont {Stern}, \citenamefont
  {Harrington}, \citenamefont {Lita}, \citenamefont {Gerrits}, \citenamefont
  {Vayshenker}, \citenamefont {Baek}, \citenamefont {Shaw}, \citenamefont
  {Mirin},\ and\ \citenamefont {Nam}}]{Marsili2013}%
  \BibitemOpen
  \bibfield  {author} {\bibinfo {author} {\bibfnamefont {F.}~\bibnamefont
  {Marsili}}, \bibinfo {author} {\bibfnamefont {V.~B.}\ \bibnamefont {Verma}},
  \bibinfo {author} {\bibfnamefont {J.~A.}\ \bibnamefont {Stern}}, \bibinfo
  {author} {\bibfnamefont {S.}~\bibnamefont {Harrington}}, \bibinfo {author}
  {\bibfnamefont {A.~E.}\ \bibnamefont {Lita}}, \bibinfo {author}
  {\bibfnamefont {T.}~\bibnamefont {Gerrits}}, \bibinfo {author} {\bibfnamefont
  {I.}~\bibnamefont {Vayshenker}}, \bibinfo {author} {\bibfnamefont
  {B.}~\bibnamefont {Baek}}, \bibinfo {author} {\bibfnamefont {M.~D.}\
  \bibnamefont {Shaw}}, \bibinfo {author} {\bibfnamefont {R.~P.}\ \bibnamefont
  {Mirin}}, \ and\ \bibinfo {author} {\bibfnamefont {S.~W.}\ \bibnamefont
  {Nam}},\ }\href@noop {} {\bibfield  {journal} {\bibinfo  {journal} {Nature
  Photon.}\ }\textbf {\bibinfo {volume} {7}},\ \bibinfo {pages} {210} (\bibinfo
  {year} {2013})}\BibitemShut {NoStop}%
\bibitem [{\citenamefont {Safavi-Naeini}\ \emph {et~al.}(2013)\citenamefont
  {Safavi-Naeini}, \citenamefont {Chan}, \citenamefont {Hill}, \citenamefont
  {Gr\"oblacher}, \citenamefont {Miao}, \citenamefont {Chen}, \citenamefont
  {Aspelmeyer},\ and\ \citenamefont {Painter}}]{Safavi-Naeini2013a}%
  \BibitemOpen
  \bibfield  {author} {\bibinfo {author} {\bibfnamefont {A.~H.}\ \bibnamefont
  {Safavi-Naeini}}, \bibinfo {author} {\bibfnamefont {J.}~\bibnamefont {Chan}},
  \bibinfo {author} {\bibfnamefont {J.~T.}\ \bibnamefont {Hill}}, \bibinfo
  {author} {\bibfnamefont {S.}~\bibnamefont {Gr\"oblacher}}, \bibinfo {author}
  {\bibfnamefont {H.}~\bibnamefont {Miao}}, \bibinfo {author} {\bibfnamefont
  {Y.}~\bibnamefont {Chen}}, \bibinfo {author} {\bibfnamefont {M.}~\bibnamefont
  {Aspelmeyer}}, \ and\ \bibinfo {author} {\bibfnamefont {O.}~\bibnamefont
  {Painter}},\ }\href@noop {} {\bibfield  {journal} {\bibinfo  {journal} {New
  J.\ Phys.}\ }\textbf {\bibinfo {volume} {15}},\ \bibinfo {pages} {035007}
  (\bibinfo {year} {2013})}\BibitemShut {NoStop}%
\bibitem [{\citenamefont {Poggio}\ \emph {et~al.}(2007)\citenamefont {Poggio},
  \citenamefont {Degen}, \citenamefont {Mamin},\ and\ \citenamefont
  {Rugar}}]{Poggio2007}%
  \BibitemOpen
  \bibfield  {author} {\bibinfo {author} {\bibfnamefont {M.}~\bibnamefont
  {Poggio}}, \bibinfo {author} {\bibfnamefont {C.~L.}\ \bibnamefont {Degen}},
  \bibinfo {author} {\bibfnamefont {H.~J.}\ \bibnamefont {Mamin}}, \ and\
  \bibinfo {author} {\bibfnamefont {D.}~\bibnamefont {Rugar}},\ }\href@noop {}
  {\bibfield  {journal} {\bibinfo  {journal} {Phys.\ Rev.\ Lett.}\ }\textbf
  {\bibinfo {volume} {99}},\ \bibinfo {pages} {017201} (\bibinfo {year}
  {2007})}\BibitemShut {NoStop}%
\bibitem [{\citenamefont {Rocheleau}\ \emph {et~al.}(2010)\citenamefont
  {Rocheleau}, \citenamefont {Ndukum}, \citenamefont {Macklin}, \citenamefont
  {Hertzberg}, \citenamefont {Clerk},\ and\ \citenamefont
  {Schwab}}]{Rocheleau2010}%
  \BibitemOpen
  \bibfield  {author} {\bibinfo {author} {\bibfnamefont {T.}~\bibnamefont
  {Rocheleau}}, \bibinfo {author} {\bibfnamefont {T.}~\bibnamefont {Ndukum}},
  \bibinfo {author} {\bibfnamefont {C.}~\bibnamefont {Macklin}}, \bibinfo
  {author} {\bibfnamefont {J.~B.}\ \bibnamefont {Hertzberg}}, \bibinfo {author}
  {\bibfnamefont {A.~A.}\ \bibnamefont {Clerk}}, \ and\ \bibinfo {author}
  {\bibfnamefont {K.~C.}\ \bibnamefont {Schwab}},\ }\href@noop {} {\bibfield
  {journal} {\bibinfo  {journal} {Nature}\ }\textbf {\bibinfo {volume} {463}},\
  \bibinfo {pages} {72} (\bibinfo {year} {2010})}\BibitemShut {NoStop}%
\bibitem [{\citenamefont {Hofer}\ \emph {et~al.}(2011)\citenamefont {Hofer},
  \citenamefont {Wieczorek}, \citenamefont {Aspelmeyer},\ and\ \citenamefont
  {Hammerer}}]{Hofer2011}%
  \BibitemOpen
  \bibfield  {author} {\bibinfo {author} {\bibfnamefont {S.~G.}\ \bibnamefont
  {Hofer}}, \bibinfo {author} {\bibfnamefont {W.}~\bibnamefont {Wieczorek}},
  \bibinfo {author} {\bibfnamefont {M.}~\bibnamefont {Aspelmeyer}}, \ and\
  \bibinfo {author} {\bibfnamefont {K.}~\bibnamefont {Hammerer}},\ }\href@noop
  {} {\bibfield  {journal} {\bibinfo  {journal} {Phys. Rev. A}\ }\textbf
  {\bibinfo {volume} {84}},\ \bibinfo {pages} {052327} (\bibinfo {year}
  {2011})}\BibitemShut {NoStop}%
\bibitem [{\citenamefont {Nielsen}\ and\ \citenamefont
  {Chuang}(2000)}]{Nielsen2000}%
  \BibitemOpen
  \bibfield  {author} {\bibinfo {author} {\bibfnamefont {M.}~\bibnamefont
  {Nielsen}}\ and\ \bibinfo {author} {\bibfnamefont {I.}~\bibnamefont
  {Chuang}},\ }\href@noop {} {\emph {\bibinfo {title} {Quantum Computation and
  Quantum Information}}}\ (\bibinfo  {publisher} {Cambridge University Press},\
  \bibinfo {year} {2000})\BibitemShut {NoStop}%
\bibitem [{\citenamefont {Duan}\ \emph {et~al.}(2001)\citenamefont {Duan},
  \citenamefont {Lukin}, \citenamefont {Cirac},\ and\ \citenamefont
  {Zoller}}]{Duan2001}%
  \BibitemOpen
  \bibfield  {author} {\bibinfo {author} {\bibfnamefont {L.~M.}\ \bibnamefont
  {Duan}}, \bibinfo {author} {\bibfnamefont {M.~D.}\ \bibnamefont {Lukin}},
  \bibinfo {author} {\bibfnamefont {J.~I.}\ \bibnamefont {Cirac}}, \ and\
  \bibinfo {author} {\bibfnamefont {P.}~\bibnamefont {Zoller}},\ }\href@noop {}
  {\bibfield  {journal} {\bibinfo  {journal} {Nature}\ }\textbf {\bibinfo
  {volume} {414}},\ \bibinfo {pages} {413} (\bibinfo {year}
  {2001})}\BibitemShut {NoStop}%
\bibitem [{\citenamefont {Sangouard}\ \emph {et~al.}(2011)\citenamefont
  {Sangouard}, \citenamefont {Simon}, \citenamefont {de~Riedmatten},\ and\
  \citenamefont {Nicolas}}]{Sangouard2011}%
  \BibitemOpen
  \bibfield  {author} {\bibinfo {author} {\bibfnamefont {N.}~\bibnamefont
  {Sangouard}}, \bibinfo {author} {\bibfnamefont {C.}~\bibnamefont {Simon}},
  \bibinfo {author} {\bibfnamefont {H.}~\bibnamefont {de~Riedmatten}}, \ and\
  \bibinfo {author} {\bibnamefont {Nicolas}},\ }\href@noop {} {\bibfield
  {journal} {\bibinfo  {journal} {Rev. Mod. Phys.}\ }\textbf {\bibinfo {volume}
  {83}},\ \bibinfo {pages} {33} (\bibinfo {year} {2011})}\BibitemShut {NoStop}%
\end{thebibliography}
\end{document}